\documentclass[aps, prb, twocolumn, superscriptaddress, longbibliography]{revtex4-2}
\usepackage{amsmath,amssymb}
\usepackage{bm}
\usepackage{color}
\usepackage{graphicx}
\usepackage[
 colorlinks=true,
 citecolor=blue,
 linkcolor=blue
]{hyperref}

\let\Re\relax
\let\Im\relax
\DeclareMathOperator{\Re}{Re}
\DeclareMathOperator{\Im}{Im}

\graphicspath{{./Figure/}}
\begin{document}
\title{Theory of Magnetic Excitations \\ in the Heavy-Fermion Spin-Triplet Superconductor UTe$_2$}
\author{Koki Shimura}
\email[]{k-shimura@g.ecc.u-tokyo.ac.jp}
\affiliation{%
 Department of Physics, The University of Tokyo,
 7-3-1 Hongo, Tokyo 113-0033, Japan
}

\author{Shuntaro Sumita}
\affiliation{%
 Department of Basic Science, The University of Tokyo,
 3-8-1 Komaba, Tokyo 153-8902, Japan
}
\affiliation{%
 RIKEN Center for Emergent Matter Science,
 Wako, Saitama 351-0198, Japan
}

\author{Yusuke Kato}
\affiliation{%
 Department of Physics, The University of Tokyo,
 7-3-1 Hongo, Tokyo 113-0033, Japan
}
\affiliation{%
 Department of Basic Science, The University of Tokyo,
 3-8-1 Komaba, Tokyo 153-8902, Japan
}
\affiliation{%
 Quantum Research Center for Chirality, Institute for Molecular Science,
 Okazaki, Aichi 444-8585, Japan
}

\date{\today}

\begin{abstract}
We study the dynamical spin response of UTe$_2$ by using a mixed-dimensional periodic Anderson model. Within the BCS-RPA formalism, we examine how the $f$-orbital character of the quasiparticles affects magnetic excitations in both the normal and superconducting (SC) states. In the normal state, finite mixing between localized $f$ electrons and conduction electrons produces a hybridization gap and enhances the spin response at $\mathbf{Q}_{\mathrm{Y}} = (0,\pi,0)$, indicating that the magnetic excitation originates from particle--hole scattering across the hybridization gap. In the SC state, we compare four odd-parity irreducible representations, $A_u$, $B_{1u}$, $B_{2u}$, and $B_{3u}$, for the spin-triplet order parameter. We find that, for the component of the spin susceptibility parallel to the $\mathbf{d}$ vector, a pronounced superconductivity-induced spin resonance appears at $\mathbf{Q}_{\mathrm{Y}}$ only in the $B_{2u}$ state. This behavior arises because the $B_{2u}$ order parameter remains finite and changes its sign between the relevant $f$-electron-dominated Fermi-surface regions connected by $\mathbf{Q}_\mathrm{Y}$ near $k_z=\pi$. The sign-change criterion is applicable to multiband superconductors in three-dimensional heavy-fermion systems, in the presence of (i) low-dimensional portions of the Fermi surface connected by a nesting vector, (ii) the dominance of the $f$-electron character, and (iii) the finite amplitude of SC gap on the portions.
\end{abstract}

\maketitle

\section{Introduction}
Inelastic neutron scattering (INS) is a powerful experimental probe of superconducting (SC) gap symmetry through the dynamical spin susceptibility, and has played an important role in studies of unconventional superconductors, including high-$T_{\mathrm{c}}$ cuprates~\cite{Mook1993PRL,Mook1998Nature,Dai1998PRL,Dai1999Science,Fong1999Nature,He2002Science,Reznik2004PRL,Pailhes2004PRL,Vignolle2007NatPhys,Tranquada2004Nature,Keimer2015Nature,Fujita2012JPSJ} and iron-based superconductors~\cite{Xie2018PRLMar,Xie2018PRLJun,Wang2016,Zhang2013,Steffens2013,Li2009,Chi2009,Shamoto2010,Christianson2008}. In these systems, a resonant enhancement of the spin response in the SC state has been discussed in relation to the symmetry of the SC order parameter. In particular, in high-$T_{\mathrm{c}}$ cuprates, the spin resonance has been discussed mainly within effective single-band models for the CuO$_2$ plane on a square lattice, and its origin is attributed to a sign change in the SC order parameter between Fermi-surface regions connected by the antiferromagnetic (AFM) nesting vector $\mathbf{Q}$, i.e., $\Delta^{\mathbf{k}+\mathbf{Q}} = -\Delta^{\mathbf{k}}$~\cite{Bulut1996,Morr1998,Abanov1999,Dahm1998,Norman1998,Eschrig2006}. In the following, we refer to this modification of the magnetic excitation spectrum caused by the opening of the SC gap as the SC feedback effect~\cite{Thalmeier2010}.

In heavy-fermion superconductors, changes in magnetic excitations below the SC transition temperature have also been investigated by INS. In CeCoIn$_5$~\cite{Stock2008PRL,Chubukov2008PRL,Akbari2012PRB}, CeCu$_2$Si$_2$~\cite{Stockert2011NatPhys,Eremin2008PRL}, and UPd$_2$Al$_3$~\cite{Metoki1998PRL,Bernhoeft1998PRL,Sato2001Nature,Chang2007PRB,McHale2004PRB}, spin resonances have been observed in the SC state, and their relationship to the SC order parameter and pairing mechanism has been discussed. More recently, UTe$_2$ has attracted considerable attention as a promising candidate for spin-triplet superconductivity, and its magnetic excitations have also been studied by INS~\cite{Duan2020,Halloran2025,Butch2022,Knafo2021PRB}. Interestingly, although spin-triplet pairing is generally driven by ferromagnetic fluctuations, the coexistence of ferromagnetic and AFM fluctuations has been reported in UTe$_2$~\cite{Ambika2022}. Other INS experiments have revealed incommensurate AFM fluctuations in the normal state near $\mathbf{q}_\mathrm{exp}=(0, 0.57, 0)\,\mathrm{r.l.u.}$, which corresponds to $(0, 1.14\pi, 0)$ in our convention, with a maximum intensity around $\omega \sim 4\,\mathrm{meV}$~\cite{Duan2020,Halloran2025}. The energy scale of the 4 meV magnetic excitation is close to that of the Kondo hybridization gap observed by scanning tunneling spectroscopy~\cite{Jiao2020}, suggesting that the normal-state magnetic excitation is related to the hybridized electronic structure. Upon entering the SC state, a low-energy peak develops near $\omega \sim 1\,\mathrm{meV}$ at a similar wave vector, which has been interpreted as a spin resonance arising from the SC feedback effect~\cite{Duan2021,Raymond2021}.

However, in heavy-fermion superconductors, it is generally difficult to apply the single-band picture that has been adopted for cuprates~\cite{Bulut1996,Morr1998,Abanov1999,Dahm1998,Norman1998,Eschrig2006}. In heavy-fermion systems, localized $f$ electrons hybridize with conduction electrons, leading to the formation of heavy quasiparticle bands near the Fermi energy. This hybridization can enhance the dynamical spin response at finite frequencies even in the normal state, in contrast to the SC feedback effects. Indeed, in YbB$_{12}$ and CeB$_6$, hybridized heavy quasiparticle bands can give rise to pronounced dynamical spin responses and collective spin-exciton modes~\cite{Akbari2009,Akbari2012PRL}. Related spin-exciton scenarios have also been discussed for heavy-fermion superconductors such as Ce$M$In$_5$ ($M=$ Co, Rh, and Ir)~\cite{Akbari2012PRB}. Therefore, understanding the magnetic excitations observed in UTe$_2$ requires a microscopic analysis of the dynamical spin response that takes into account both the hybridized electronic structure in the normal state and the feedback effect in the SC state.

UTe$_2$ is also of particular interest for the symmetry of the SC order parameter, which is still under debate. Its upper critical field far exceeds the Pauli limit~\cite{Ran2019,Aoki2019, Aoki2022_JPCM}, and the decrease in the Knight shift below the SC transition temperature is smaller than that expected for conventional spin-singlet pairing~\cite{Matsumura2023}. Thermal conductivity, specific heat, and penetration-depth measurements suggest the existence of point nodes~\cite{Metz2019,Hayes2025,Kittaka2020,Ishihara2023}, whereas another thermal-conductivity measurement and NMR Knight-shift studies on high-quality samples have been interpreted as supporting a fully gapped or pseudo-point-nodal SC state~\cite{Suetsugu2024,Matsumura2023,Hosoi2026}. These observations have motivated proposals of various odd-parity SC states~\cite{Ran2019,Aoki2019,Aoki2022_JPCM,Metz2019,Kittaka2020,Suetsugu2024}, but no consensus on the SC symmetry has yet been reached.

Since the SC feedback effect is sensitive to the momentum dependence of the SC order parameter, comparing the magnetic responses can help us distinguish among the candidate states. However, microscopic theoretical studies of the dynamical spin response in the SC states of UTe$_2$ remain limited. Focusing on the $B_{3u}$ irreducible representation, Chen et al.~\cite{Chen2021} showed that multiorbital spin-triplet pairing can give rise to an AFM spin resonance. However, the hybridization between localized $f$ electrons and conduction electrons was not taken into account. It therefore remains unclear how the hybridization gap modifies the magnetic response in the normal state and how this response can be distinguished from the superconductivity-induced enhancement.

The Fermi surface of UTe$_2$ and the $f$-electron character of its low-energy quasiparticles have been studied through both experiments~\cite{Fujimori2019,Miao2020,Broyles2023,Weinberger2024,Eaton2024} and first-principles calculations~\cite{Xu2019,Ishizuka2019,Shick2021,Shishidou2021}. The mixed-dimensional periodic Anderson model proposed by Hakuno et al. incorporates the key features of the electronic structure suggested by these studies and reproduces the momentum structure of the AFM fluctuations observed by INS~\cite{Hakuno2024}. Therefore, this model provides a suitable microscopic framework for investigating magnetic excitations arising from hybridization in the normal state and spin resonances induced by superconductivity.

In this work, we aim to clarify how the Fermi-surface geometry, the orbital character of the low-energy quasiparticles, and the momentum dependence of the SC order parameter affect the magnetic response. We analyze the dynamical spin susceptibility of UTe$_2$ within the BCS-RPA formalism~\cite{Bulut1996,Morr1998,Dahm1998,Eschrig2006,
Akbari2012PRB,Eremin2008PRL,Yoshikawa1999,Korshunov2008,
Maier2008,Maier2009,Onari2010,Chen2022}, based on the mixed-dimensional periodic Anderson model~\cite{Hakuno2024}. In the normal state, we show that finite hybridization opens a gap in the low-energy electronic structure and enhances the dynamical spin susceptibility at the AFM wave vector $\mathbf{Q}_{\mathrm{Y}}=(0,\pi,0)$. In the SC state, on the other hand, we consider all four odd-parity irreducible representations, $A_u$, $B_{1u}$, $B_{2u}$, and $B_{3u}$, of the $D_{2h}$ point group. We find that, for the longitudinal spin response parallel to the $\mathbf d$ vector considered here, a pronounced spin resonance due to the SC feedback effect appears only for the $B_{2u}$ order parameter. This result arises because the SC order parameter in the $B_{2u}$ state remains finite and takes opposite signs on the $f$-electron-dominated Fermi-surface regions near $k_z=\pi$ that are connected by $\mathbf{Q}_{\mathrm{Y}}$. Thus, although UTe$_2$ is intrinsically multiband, the magnetic response in the SC state can be understood by applying the sign-reversal criterion to the $f$-electron-dominated quasiparticle states that mainly contribute to the magnetic response. Our results provide a microscopic basis for interpreting INS spectra in UTe$_2$ and offer insight into the relationship between magnetic excitations and SC pairing symmetry in multiband heavy-fermion systems.

This paper is organized as follows. In Sec.~\ref{sec:model}, we introduce the mixed-dimensional periodic Anderson model for UTe$_2$~\cite{Hakuno2024} and formulate the dynamical spin susceptibility within the BCS-RPA formalism. In Sec.~\ref{sec:results}, we present the magnetic responses in both the normal and SC states. We first discuss the magnetic excitation induced by the hybridization gap in the normal state. We then examine the SC state for the $A_u$, $B_{1u}$, $B_{2u}$, and $B_{3u}$ order parameters and clarify the condition for the superconductivity-induced spin resonance. Section~\ref{sec:summary} summarizes the present study. Throughout this paper, we set $\hbar=1$.

\section{Model and Method}
\label{sec:model}

In this section, we introduce the theoretical framework used to analyze magnetic excitations in UTe$_2$.
We first describe the mixed-dimensional periodic Anderson model and the SC order parameters.
Next, we formulate the dynamical spin susceptibility within the BCS-RPA formalism.

\subsection{Mixed-dimensional periodic Anderson model}
\label{sec:model-MA}
We use the mixed-dimensional periodic Anderson model introduced in Ref.~\cite{Hakuno2024},
\begin{equation}
 H = H_0 + H_U,
 \label{eq:H}
\end{equation}
where
\begin{align}
 H_0 &= \sum_{\mathbf{k} m \sigma} \left(\varepsilon_m^{\mathbf{k}}-\mu\right) a^\dagger_{m \mathbf{k} \sigma} a_{m \mathbf{k} \sigma} \notag \\
 &\quad + \sum_{\mathbf{k} m l\sigma}\left(V_{ml} a^\dagger_{m \mathbf{k} \sigma} a_{l \mathbf{k} \sigma} + \mathrm{H.c.}\right),
 \label{eq:H_0} \\
 H_U &= U \sum_i n_{fi\uparrow} n_{fi\downarrow}.
 \label{eq:H_U}
\end{align}
Here, $H_0$ describes the non-interacting part of a three-orbital tight-binding model defined on an effective orthorhombic lattice with $D_{2h}$ point-group symmetry. The $x$, $y$, and $z$ directions correspond to the crystallographic $a$, $b$, and $c$ axes, respectively. The operator $a^\dagger_{m\mathbf{k}\sigma}$ ($a_{m\mathbf{k}\sigma}$) creates (annihilates) an electron in orbital $m$ with momentum $\mathbf{k}$ and spin $\sigma$. The first term in Eq.~(\ref{eq:H}) represents the kinetic energy, where $m=f,d,p$ labels the three orbitals. The dispersion relations are given by
\begin{subequations}
 \begin{align}
  \varepsilon_{f}^{\mathbf{k}} &= -2t_{fx}\cos k_{x} - 2t_{fy}\cos k_{y} \notag \\
  &\quad + 2t_{fz}(\cos k_{z}+1) + \varepsilon^0_f, 
  \label{eq:dispersion_f} \\
  \varepsilon_{d}^{\mathbf{k}} &= -2t_{dx}\cos k_{x} - 2t_{dy}\cos k_{y} + \varepsilon^0_d,
  \label{eq:dispersion_d} \\
  \varepsilon_{p}^{\mathbf{k}} &= 2t_{px}\cos k_{x} + 2t_{py}\cos k_{y} + \varepsilon^0_p,
  \label{eq:dispersion_p}
 \end{align}
\end{subequations}
where the crystal fields are denoted by $\varepsilon^0_m$. The second term describes the hybridization $V_{ml}$ between $m$ and $l$ orbitals, with the orbital indices $(m,l) = (f,d), (f,p), (d,p)$. Unless noted otherwise, all calculations are performed with the parameter set $(t_{fx},\,t_{fy},\,t_{fz},\,t_{dx},\,t_{dy},\,\varepsilon^0_d,\,t_{px},\,t_{py},\,\varepsilon^0_p,\,V_{fd} ,\,V_{dp},\,\mu)
= (0.08,\,0.035,\,0.1,\,0.5,\,0,\,0,\,0,\,1,\,0,\,0.05,\,0.05,\,-0.1)$. This parameter set follows Ref.~\cite{Hakuno2024}, with $t_{py}=1$ taken as the unit of energy. Unless otherwise specified, we set $\varepsilon_f^0=0.05$ and $V_{fp}=0.05$. We vary $V_{fp}$ as a tuning parameter to examine changes in the geometry of the Fermi surface and nesting properties. The resulting band structure captures qualitative features of the Fermi surfaces proposed for UTe$_2$~\cite{Fujimori2019,Miao2020,Broyles2023,Weinberger2024,Eaton2024,Halloran2025,Xu2019,Ishizuka2019,Shick2021,Shishidou2021} and produces enhanced AFM fluctuations near $\mathbf{Q}_{\mathrm{Y}}=(0,\pi,0)$, consistent with INS observations~\cite{Duan2020,Butch2022} in the normal state. Equation~\eqref{eq:H_U} represents the on-site Coulomb interaction among the $f$ electrons, where $n_{fi\sigma}$ ($\sigma = \, \uparrow, \downarrow$) is the number operator for the $f$ electron on site $i$ with spin $\sigma$.

Note that the Hamiltonian in Eq.~\eqref{eq:H} neglects the spin--orbit coupling (SOC) and the sublattice degrees of freedom. Although a more realistic model incorporating both SOC and the sublattice structure has recently been proposed~\cite{Hakuno2026}, the present model captures the hybridization-induced reconstruction of the band structure and the redistribution of $f$-electron weight over the low-dimensional Fermi surface, as well as the momentum dependence of the SC order parameter, which are the essential ingredients needed to address the present analysis. We therefore regard this SOC-free model as a minimal framework for isolating the fundamental momentum-space mechanisms governing magnetic excitations and spin resonances.

\begin{figure}[tbp]
 \centering
 \includegraphics[width=\linewidth]{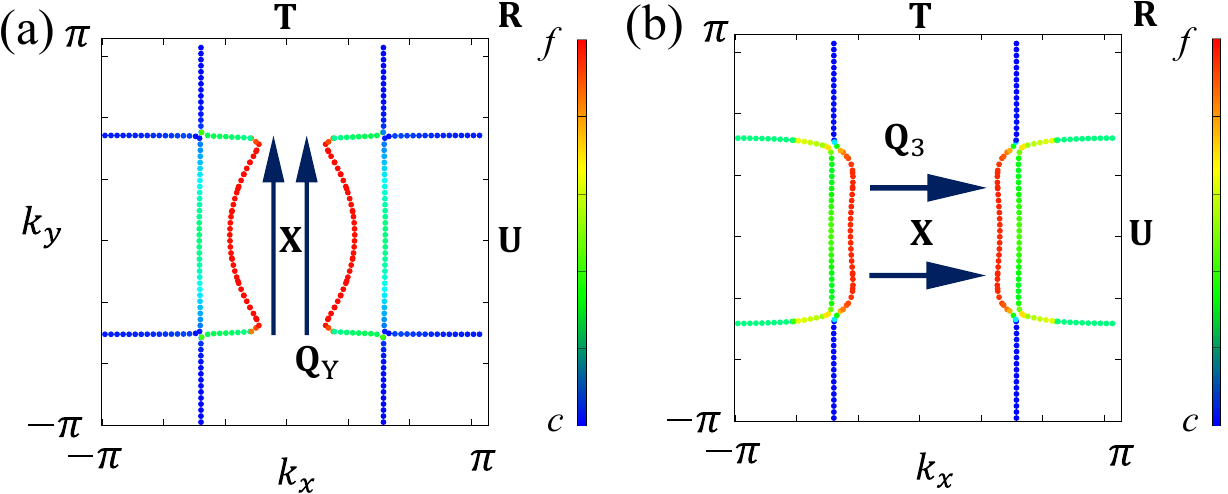}
 \caption{Fermi surfaces at $k_z=\pi$ and real part of the static bare susceptibility $\Re\,\chi_0(\mathbf{q}, \omega = 0)$ with $\varepsilon^0_f=0.05$ for (a) $V_{fp}=0.05$ and (b) $V_{fp}=0.26$~\cite{Hakuno2024}. Here, nesting vectors are illustrated by arrows, and the color represents the $f$-orbital weight of the normal-state band eigenstate on the Fermi surface; red and blue correspond to predominantly $f$-electron and conduction-electron character, respectively.}
 \label{fig:bandFermi}
\end{figure}

Before introducing the SC state, we briefly review the properties of the Fermi surface in the normal state for the present model, following Ref.~\cite{Hakuno2024}. Figure~\ref{fig:bandFermi} shows the Fermi surfaces on the $k_z=\pi$ plane for two representative values of the hybridization: (a) $V_{fp}=0.05$ and (b) $V_{fp}=0.26$, together with the dominant nesting vectors $\mathbf{Q}_\mathrm{Y} = (0, \pi, 0)$ and $\mathbf{Q}_\mathrm{3} = (0.69\pi, 0, 0)$. For $V_{fp}=0.05$, the Fermi surfaces on the $k_z=\pi$ plane consist of sheets with a quasi-two-dimensional shape and a pocket dominated by $f$ electrons. The quasiparticle states on the sheets contain substantial contributions from conduction electrons. This structure is consistent with previous reports from ARPES~\cite{Miao2020} and measurements of quantum oscillations~\cite{Broyles2023}. LDA+U calculations that include an intermediate Coulomb interaction also reproduce the pocket dominated by $f$ electrons~\cite{Ishizuka2019} with the nesting vector $\mathbf{Q}_{\mathrm{Y}}$. As the hybridization $V_{fp}$ increases, the geometry of the Fermi surface changes. For $V_{fp}=0.26$, the dominant nesting vector shifts toward the $q_x$ direction and becomes $\mathbf{Q}_3$. Related changes in the momentum structure associated with modifications of the electronic structure and the $f$-orbital weight have also been discussed in multiorbital studies of UTe$_2$ under pressure~\cite{Ishizuka2021,Shimizu2026}.
  
Next, to describe the SC state, we introduce the Bogoliubov--de Gennes (BdG) Hamiltonian
\begin{equation}
 H_{\mathrm{BdG}} = \sum_{\mathbf{k}} \mathbf{\Psi}_{\mathbf{k}}^\dagger \hat{H}_{\mathrm{BdG}}(\mathbf{k}) \mathbf{\Psi}_{\mathbf{k}},
\end{equation}
by using the Nambu spinor
\begin{equation}
 \mathbf{\Psi}_{\mathbf{k}}
 =
 \left(
 a_{f,\mathbf{k},\uparrow},
 a_{d,\mathbf{k},\uparrow},
 a_{p,\mathbf{k},\uparrow},
 a_{f,-\mathbf{k},\downarrow}^{\dagger},
 a_{d,-\mathbf{k},\downarrow}^{\dagger},
 a_{p,-\mathbf{k},\downarrow}^{\dagger}
 \right)^{\mathrm{T}}.
 \label{NambuBasis}
\end{equation}
The BdG Hamiltonian matrix is given by
\begin{equation}
 \hat{H}_{\mathrm{BdG}}(\mathbf{k}) =
 \begin{pmatrix}
  \underline{H}_0(\mathbf{k}) & \underline{\Delta}(\mathbf{k}) \\
  \underline{\Delta}^\dagger(\mathbf{k}) & -\underline{H}_0^{\mathrm{T}}(-\mathbf{k})
 \end{pmatrix}.
 \label{eq:H_BdG}
\end{equation}
Here, $\underline{\Delta}(\mathbf{k})$ is the $3 \times 3$ SC order parameter matrix in the orbital space, and $\mathrm{T}$ means the transpose. Since the low-energy quasiparticle states contributing to the magnetic response have a large $f$-orbital weight, we assume that pairing occurs predominantly on the $f$ orbitals. Then the normal-state Hamiltonian matrix and the SC order parameter matrix are given by
\begin{align}
 & \underline{H}_0(\mathbf{k}) =
 \begin{pmatrix}
  \varepsilon_{f}^{\mathbf{k}}-\mu & V_{fd} & V_{fp} \\
  V_{fd} & \varepsilon_{d}^{\mathbf{k}}-\mu & V_{dp} \\
  V_{fp} & V_{dp} & \varepsilon_{p}^{\mathbf{k}}-\mu
 \end{pmatrix},
 \\
 & \underline{\Delta}(\mathbf{k}) =
 \begin{pmatrix}
  \Delta^{\mathbf{k}} & 0 & 0 \\
  0 & 0 & 0 \\
  0 & 0 & 0
 \end{pmatrix}.
\end{align}
The BdG Hamiltonian $\hat{H}_{\mathrm{BdG}}(\mathbf{k})$ is diagonalized by using a unitary matrix $\hat{W}(\mathbf{k})$ as
\begin{align}
&\hat{W}^{\dagger}(\mathbf{k})
\hat{H}_{\mathrm{BdG}}(\mathbf{k})
\hat{W}(\mathbf{k}) \notag \\
&=
\mathrm{diag}
\Bigl(
E_1^{\mathbf{k}},
E_2^{\mathbf{k}},
E_3^{\mathbf{k}},
E_4^{\mathbf{k}},
E_5^{\mathbf{k}},
E_6^{\mathbf{k}}
\Bigr),
\label{eq:H_BdG_diagonalization}
\end{align}
where $E_j^{\mathbf{k}}$ represents the $j$th energy band for Bogoliubov quasiparticles.

We next specify the explicit form of the SC order parameter $\Delta^{\mathbf{k}}$. We do not determine the SC order parameter self-consistently, but instead assume a representative basis function for each irreducible representation. In the absence of SOC, the orientation of the $\mathbf{d}$ vector is arbitrary in spin space. We therefore choose $\mathbf{d}(\mathbf{k})=\hat{\mathbf{z}}\Delta^{\mathbf{k}}$ throughout this paper. Under the point group $D_{2h}$, the odd-parity irreducible representations for spin-triplet pairing are $A_u$, $B_{1u}$, $B_{2u}$, and $B_{3u}$, each corresponding to a different gap function $\Delta^{\mathbf{k}}$ with different nodal structures. Table~\ref{tab:gap_sym} summarizes the odd-parity irreducible representations of the $D_{2h}$ point group and their corresponding basis functions. The $B_{1u}$, $B_{2u}$, and $B_{3u}$ representations correspond to $p$-wave superconductivity. Note that the simplest basis function of $A_u$ is $\sin k_x \sin k_y \sin k_z$, which represents a nodal $f$-wave state.
This is the consequence of neglecting SOC in our model; indeed, in strongly spin--orbit coupled systems, a fully gapped $p$-wave order parameter belongs to the $A_u$ representation~\cite{Sigrist1991}.
As mentioned above, however, one of our purposes is to understand the interplay among orbital hybridization, SC feedback effects, and spin responses based on the simple model.
Therefore, we discuss only the SOC-free case as a first step; we leave the effects of SOC on the spin responses for future study.

\begin{table}[t]
 \caption{Basis functions of the SC order parameter for all odd-parity irreducible representations of the point group $D_{2h}$. $\Delta_0$ is a gap amplitude.}
 \label{tab:gap_sym}
 \centering
 \begin{tabular}{cc}
  \hline\hline
  Symmetry & $\Delta^{\mathbf{k}}$ \\
  \hline
  $A_u$ & $\Delta_0 \sin k_x \sin k_y \sin k_z$ \\
  $B_{1u}$ & $\Delta_0 \sin k_z$ \\
  $B_{2u}$ & $\Delta_0 \sin k_y$ \\
  $B_{3u}$ & $\Delta_0 \sin k_x$ \\
  \hline\hline
 \end{tabular}
\end{table}

\subsection{Dynamical spin susceptibility and BCS-RPA formalism}
The magnetic scattering intensity observed in INS experiments is related to the imaginary part of the spin susceptibility. We define the susceptibility in the real-frequency representation as
\begin{equation}
\chi^{\mathrm{s}}(\mathbf{q},\omega)
=
\frac{2i}{N}
\int_{0}^{\infty} dt\,
e^{i\left(\omega+i\delta \right)t}
\left\langle
\left[
S_f^{z}(\mathbf{q},t),
S_f^{z}(-\mathbf{q},0)
\right]
\right\rangle .
\label{chi-s}
\end{equation}
Here, $\delta$ is a positive infinitesimal and $N$ is the number of unit cells. The operator $S_f^{z}(\mathbf{q},t)$ is the Heisenberg representation of $f$-electrons' spin operator $S_f^{z}(\mathbf{q})$, which is defined by
\begin{align}
S_f^{z}(\mathbf{q})
&=\frac{1}{2}
\sum_{\mathbf{k}}\left(
a_{f,\mathbf{k}+\mathbf{q},\uparrow}^{\dagger}
a_{f,\mathbf{k},\uparrow} - a_{f,\mathbf{k}+\mathbf{q},\downarrow}^{\dagger}
a_{f,\mathbf{k},\downarrow}\right).
\end{align}
To calculate the spin susceptibility, we adopt the BCS-RPA formalism, which has been widely used to discuss magnetic resonance in the SC state~\cite{Bulut1996,Dahm1998,Morr1998,Yoshikawa1999,Eschrig2006,
Korshunov2008,Maier2008,Maier2009,Onari2010,Eremin2008PRL,
Akbari2012PRB,Chen2022}.
Within this formalism, Eq.~\eqref{chi-s} is approximated by first evaluating the bare spin susceptibility $\chi_0(\mathbf{q},\omega)$ in the SC mean-field state and then incorporating the residual on-site interaction $U$ through the RPA ladder summation. Although the quasiparticle self-energy effects neglected in the present study may affect quantitative properties of the resonance peak, the BCS-RPA formalism captures the qualitative relationship between the symmetry of the SC order parameter and the conditions for the emergence of the resonance, which is the focus of this study.

By using this approximation, the spin susceptibility is written as
\begin{equation}
\chi^{\mathrm{s}}_{\mathrm{RPA}}(\mathbf{q}, \omega) = \frac{\chi_0(\mathbf{q},\omega)}
{1-U\chi_0(\mathbf{q},\omega)}.
\label{eqn:chi-RPA}
\end{equation}
This equation shows that the spin response is strongly enhanced when 
\begin{equation}
1-U\Re\chi_0(\mathbf{q},\omega)\simeq 0,
\qquad
\Im\chi_0(\mathbf{q},\omega)\simeq 0 .
\end{equation}
The nesting of the Fermi surface can enhance the static bare susceptibility $\Re\chi_0(\mathbf{q},0)$ in the normal state, thereby bringing the system closer to a magnetic instability. In the SC state, however, the coherence factor can suppress $\Re\chi_0(\mathbf{q},\omega)$ that would otherwise be enhanced by nesting. Therefore, the resonance condition requires the coherence factor to preserve the enhancement of $\Re\chi_0(\mathbf{q},\omega)$ arising from the nesting of the Fermi surface, allowing it to remain sufficiently large for a given $U$.

\subsubsection{Bare susceptibility}
From Eq.~\eqref{eqn:chi-RPA}, one finds that calculating the bare susceptibility $\chi_0(\mathbf{q},\omega)$ is necessary to analyze the BCS-RPA spin susceptibility $\chi^{\mathrm{s}}_{\mathrm{RPA}}(\mathbf{q},\omega)$.
For this purpose, let us introduce the Matsubara representation:
\begin{equation}
\chi_0(\mathbf{q},i\omega_m)
=\frac{2}{N}
\int_0^\beta d\tau\,
e^{i\omega_m\tau}
\left\langle
T_\tau
S_f^{z}(\mathbf{q},\tau)
S_f^{z}(-\mathbf{q},0)
\right\rangle_0,
\end{equation}
where $\beta=1/T$ is the inverse temperature, $\tau$ is the imaginary time, $\omega_m=2\pi mT$ ($m \in \mathbb{Z}$) is the bosonic Matsubara frequency, and $T_\tau$ denotes the imaginary time-ordering operator.
The statistical average $\langle \cdots \rangle_0$ is evaluated by using the eigenvectors of the BdG Hamiltonian~\eqref{eq:H_BdG}.
%, and $N$ is the number of $\mathbf{k}$-points per unit volume. 
The bare susceptibility is decomposed into the normal ($G$) and anomalous ($F$) contributions~\cite{Hara2007} as
\begin{equation}
\chi_0(\mathbf{q},i\omega_m)
=
\chi_0^{G}(\mathbf{q},i\omega_m)
+
\chi_0^{F}(\mathbf{q},i\omega_m),
\label{eq:chi0sumGandF}
\end{equation}
where
\begin{align}
\chi_0^{G}(\mathbf{q},i\omega_m)
&=
-\frac{T}{N}
\sum_{\mathbf{k},i\epsilon_n}
G_{f}(\mathbf{k}+\mathbf{q},i\epsilon_n+i\omega_m)
G_{f}(\mathbf{k},i\epsilon_n),
\label{eq:chi0_G} \\
\chi_0^{F}(\mathbf{q},i\omega_m)
&=
-\frac{T}{N}
\sum_{\mathbf{k},i\epsilon_n}
F_{f}^{\dagger}(\mathbf{k}+\mathbf{q},i\epsilon_n+i\omega_m)
F_{f}(\mathbf{k},i\epsilon_n),
\label{eq:chi0_F}
\end{align}
with $\epsilon_n=(2n+1)\pi T$ ($n \in \mathbb{Z}$) being the fermionic Matsubara frequency.  The normal and anomalous Green's functions in the $f$-orbital channel, $G_f$ and $F_f$, are respectively defined as
\begin{align}
G_{f}(\mathbf{k}, i\epsilon_n)
&=
\sum_j
\frac{
W_{fj}(\mathbf{k})W_{fj}^{*}(\mathbf{k})
}{
i\epsilon_n-E_j^{\mathbf{k}}
}, \\
F_{f}(\mathbf{k}, i\epsilon_n)
&=
\sum_j
\frac{
W_{fj}(\mathbf{k})W_{f'j}^{*}(\mathbf{k})
}{
i\epsilon_n-E_j^{\mathbf{k}}
},
\end{align}
where the subscripts $f$ and $f'$ denote the $f$-orbital component in the particle and hole channels, respectively. For $\mathbf{d}(\mathbf{k})\parallel\hat{\mathbf z}$, the anomalous contribution enters the longitudinal response and the transverse responses differently~\cite{Yakiyama2003, Morr2001, Chen2021}. In the following, we consider only the longitudinal component, which is given by Eq.~(\ref{eq:chi0sumGandF}).

The above equations contain the components of the unitary matrix $\hat{W}(\mathbf{k})$ [see Eq.~\eqref{eq:H_BdG_diagonalization}].
Note that the anomalous contributions $\chi_0^{F}$ and $F_f$ are absent in the normal state since the factor $W_{fj}(\mathbf{k})W_{f'j}^{*}(\mathbf{k})$ becomes zero.

Carrying out the summation of the Matsubara frequency in Eqs.~\eqref{eq:chi0_G} and \eqref{eq:chi0_F}, and performing the analytic continuation $i\omega_m\to\omega+i\delta$, we obtain
\begin{equation}
\chi_0(\mathbf{q},\omega)
= \frac{1}{N}
\sum_{\mathbf{k},j,l}
\mathcal{M}^{j,l}_{f}(\mathbf{k},\mathbf{q})
\frac{
n\left(E^{\mathbf{k}+\mathbf{q}}_{j}\right)-
n\left(E^{\mathbf{k}}_{l}\right)
}
{
\omega-
\left(
E^{\mathbf{k}+\mathbf{q}}_{j}-
E^{\mathbf{k}}_{l}
\right)
+i\delta
},
\label{eq:chi0_real_omega}
\end{equation}
where $n(E)$ is the Fermi--Dirac distribution function. 
We here define
\begin{equation}
\begin{aligned}
 & \mathcal{M}^{j,l}_{f}(\mathbf{k},\mathbf{q}) \notag \\
 & = \frac{1}{2}\left|W_{fj}^*(\mathbf{k}+\mathbf{q})\,W_{fl}(\mathbf{k}) + W_{f'j}^*(\mathbf{k}+\mathbf{q})\,W_{f'l}(\mathbf{k})\right|^2
\label{eq:coherence_factor}
\end{aligned}
\end{equation}
which gives a coherence factor in the multiband BdG representation, as we will discuss in Sec.~\ref{sec:coherence_factor}. 
% The first and second (third and fourth) terms originate from the normal (anomalous) Green's function.

\begin{figure}[tbp]
 \centering
 \includegraphics[width=\linewidth]{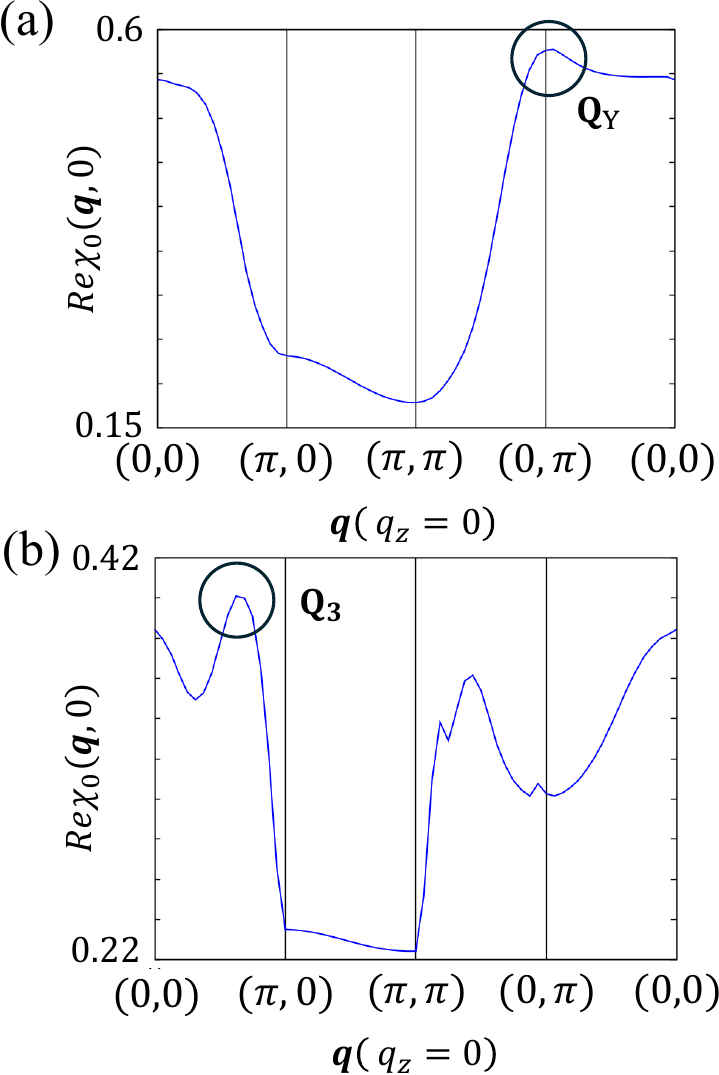}
 \caption{Real part of the static bare susceptibility $\Re\,\chi_0(\mathbf{q}, \omega = 0)$ with $\varepsilon^0_f=0.05$ for (a) $V_{fp}=0.05$ and (b) $V_{fp}=0.26$~\cite{Hakuno2024} in the normal state ($\Delta_0 = 0$). In (a), $\Re \chi_0(\mathbf{q}, 0)$ reaches its maximum at $\mathbf{Q}_{\mathrm{Y}}=(0,\pi,0)$, whereas in (b) the maximum shifts to $\mathbf{Q}_3=(0.69\pi,0,0)$.}
 \label{fig:spin_susceptibility}
\end{figure}

We next examine how the changes in the Fermi surface described in Sec.~\ref{sec:model-MA} affect the static bare spin susceptibility. Figure~\ref{fig:spin_susceptibility} shows the static spin susceptibility $\Re\chi_0(\mathbf{q},0)$ in the normal state for $V_{fp}=0.05$ and $0.26$. For $V_{fp}=0.05$, $\Re\chi_0(\mathbf{q},0)$ reaches its maximum at $\mathbf{Q}_{\mathrm{Y}}=(0,\pi,0)$, as shown in Fig.~\ref{fig:spin_susceptibility}(a). This behavior is consistent with INS experiments~\cite{Duan2020,Butch2022}. For $V_{fp}=0.26$, the maximum shifts to $\mathbf{Q}_3=(0.69\pi,0,0)$, following the change in the nesting direction of the pocket dominated by $f$ electrons [Fig.~\ref{fig:spin_susceptibility}(b)]. Thus, the wave vector that maximizes the static bare spin susceptibility depends on the hybridization through the resulting reconstruction of the Fermi surface.

\begin{figure}[tbp]
 \centering
 \includegraphics[width=\linewidth]{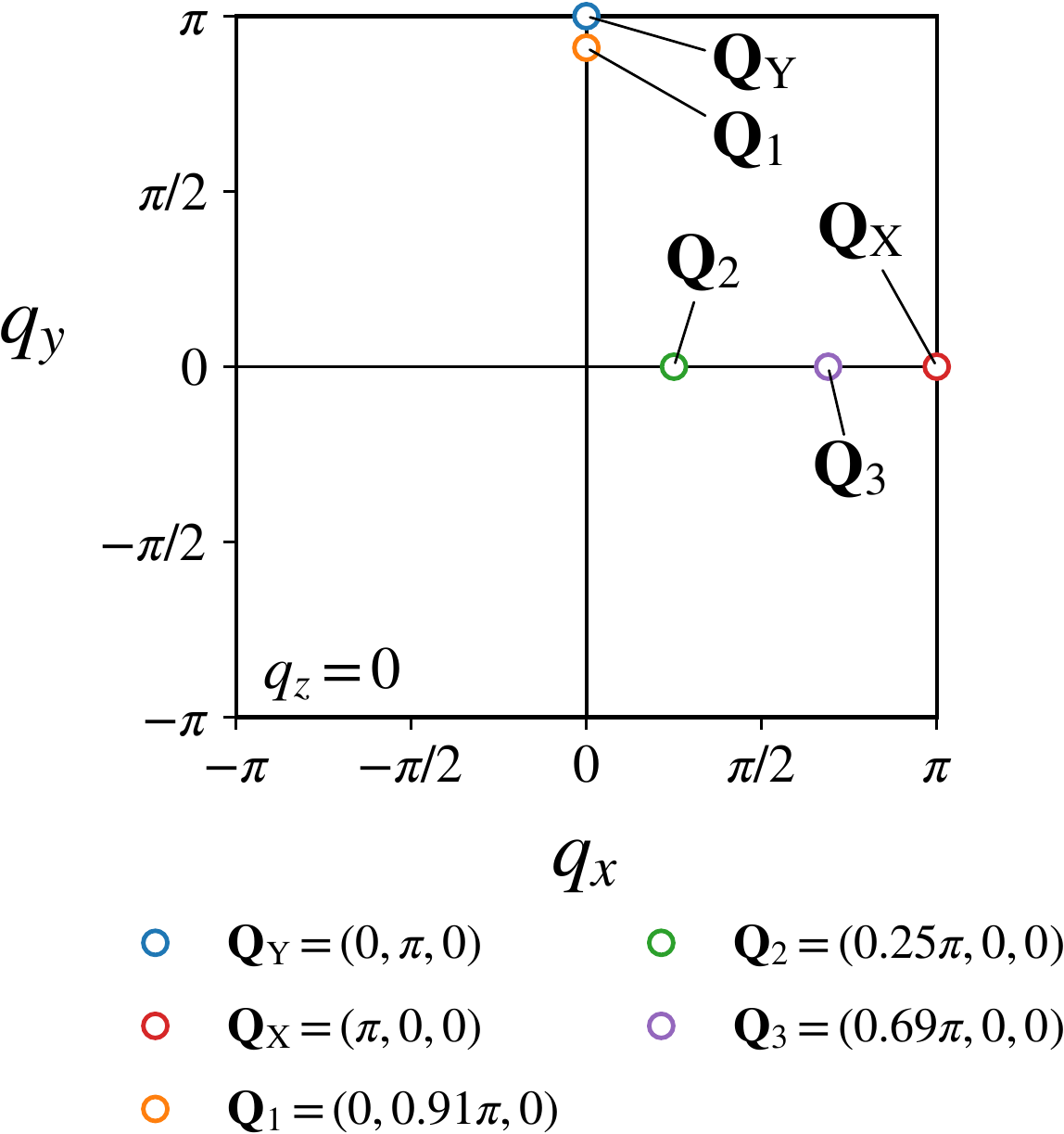}
 \caption{The positions of AFM wave vectors $\mathbf{Q}_{\mathrm{Y}}=(0,\pi,0)$, $\mathbf{Q}_{\mathrm{X}}=(\pi,0,0)$, $\mathbf{Q}_1=(0,0.91\pi,0)$, $\mathbf{Q}_2=(0.25\pi,0,0)$, and $\mathbf{Q}_3=(0.69\pi,0,0)$. For $V_{fp}=0.11$, $0.17$, and $0.26$, the static bare spin susceptibility $\chi_0(\mathbf{q},0)$ is maximized at $\mathbf{Q}_1$, $\mathbf{Q}_2$, and $\mathbf{Q}_3$, respectively.}
 \label{fig:afm_vectors}
\end{figure}

Since we are interested in the dynamical spin susceptiblity at momenta where the AFM fluctuations are enhanced, we show a few representative AFM momenta in Fig.~\ref{fig:afm_vectors}.
First, $\mathbf{Q}_{\mathrm{Y}}=(0,\pi,0)$ is the reference AFM wave vector for $V_{fp}=0.05$ [see Fig.~\ref{fig:spin_susceptibility}(a)].
The wave vector $\mathbf{Q}_{\mathrm{X}}=(\pi,0,0)$ is introduced as a momentum transfer along the $q_x$ direction for comparison with $\mathbf{Q}_{\mathrm{Y}}$.
Furthermore, $\mathbf{Q}_1$, $\mathbf{Q}_2$, and $\mathbf{Q}_3$ are the wave vectors at which the static spin susceptibility is maximized when the hybridization is $V_{fp}=0.11$, $0.17$, and $0.26$, respectively.
The vector $\mathbf{Q}_1=(0,0.91\pi,0)$ shows that AFM fluctuations along the $q_y$ direction remain dominant, while the other two vectors $\mathbf{Q}_2=(0.25\pi,0,0)$ and $\mathbf{Q}_3=(0.69\pi,0,0)$ indicate the shift of the magnetic fluctuations toward the $q_x$ direction.
In this study, we systematically investigate the spin response at these wave vectors, to disentangle the respective contributions of the $f$-$p$ hybridization $V_{fp}$ and SC feedback.

\subsubsection{Coherence factor}
\label{sec:coherence_factor}
We next discuss in detail the coherence factor defined in Eq.~\eqref{eq:coherence_factor}.
For an intuitive understanding, we first consider the orbital-decoupled limit, in which all orbital hybridization is absent: $V_{fp}=V_{dp}=V_{fd}=0$.
In this limit, band indices in the BdG spectrum are purely represented by orbital ones: $E_{j_{\pm}}^{\mathbf{k}} = \pm \sqrt{(\varepsilon_j^{\mathbf{k}})^2 + (\Delta_j^{\mathbf{k}})^2}$ ($j = f, d, p$), where $j_+$ and $j_-$ denote the positive and negative energy bands derived from orbital $j$, respectively.
Then $\mathcal{M}^{j,l}_{f}(\mathbf{k},\mathbf{q})$ is simplified as
\begin{align}
& \mathcal{M}^{j,l}_{f}(\mathbf{k},\mathbf{q}) \notag\\
&=
\begin{cases}
\displaystyle
\frac{1}{4} \left( 1 + \frac{
\varepsilon_f^{\mathbf{k}}
\varepsilon_f^{\mathbf{k}+\mathbf{q}}
+
\Delta^{\mathbf{k}}
\Delta^{\mathbf{k}+\mathbf{q}}
}{
E_{f_+}^{\mathbf{k}}
E_{f_+}^{\mathbf{k}+\mathbf{q}}
} \right)
& (j,l) = (f_\pm, f_\pm),
\\[5mm]
\displaystyle
\frac{1}{4} \left( 1 - \frac{
\varepsilon_f^{\mathbf{k}}
\varepsilon_f^{\mathbf{k}+\mathbf{q}}
+
\Delta^{\mathbf{k}}
\Delta^{\mathbf{k}+\mathbf{q}}
}{
E_{f_+}^{\mathbf{k}}
E_{f_+}^{\mathbf{k}+\mathbf{q}}
} \right)
& (j,l) = (f_\pm, f_\mp),
\\[5mm]
0
& \text{otherwise}.
\end{cases}
\label{eq:coherence_factor_f_sector}
\end{align}

Equation~\eqref{eq:coherence_factor_f_sector} corresponds to the coherence factor discussed in the previous single-band theory~\cite{Yoshikawa1999}.
Among the nonvanishing components, $\mathcal{M}^{f_\pm, f_\mp}_{f}$ describes the particle--hole quasiparticle excitations relevant to the spin resonance. On the $f$-electron Fermi surface, $\varepsilon_f^{\mathbf{k}}
\simeq
\varepsilon_f^{\mathbf{k}+\mathbf{q}}
\simeq 0$ is satisfied, and therefore the particle--hole contributions are given by
\begin{equation}
\mathcal{M}^{f_\pm, f_\mp}_{f}
\simeq
\frac{1}{4}\left(
1-
\frac{
\Delta^{\mathbf{k}}\Delta^{\mathbf{k}+\mathbf{q}}
}
{
|\Delta^{\mathbf{k}}||\Delta^{\mathbf{k}+\mathbf{q}}|
}\right).
\label{eq:coherence_factor_f_sector_p-h}
\end{equation}
Thus, when the SC order parameter changes its sign between the two quasiparticle states connected by $\mathbf{q}$, i.e., 
$\Delta^{\mathbf{k}}\Delta^{\mathbf{k}+\mathbf{q}} < 0$, 
the particle--hole scattering is enhanced. Owing to this property, for a SC order parameter with a sign reversal, $\Im\chi_0(\mathbf{q},\omega)$ rises sharply at the threshold of the particle--hole continuum, and
$\Re\chi_0(\mathbf{q},\omega)$ is enhanced through the
Kramers--Kronig relation. As a result, the denominator of the RPA spin susceptibility 
$1-U\Re\chi_0(\mathbf{q},\omega)$
becomes small, giving rise to a spin resonance.

Next, we discuss our truly multiband system by switching on the orbital hybridizations.
In this case, the normal-state Hamiltonian and the SC order parameter cannot, in general, be diagonalized simultaneously, and the coherence factor cannot be expressed by the simple form in Eq.~\eqref{eq:coherence_factor_f_sector}.
To directly evaluate the particle--hole-type quasiparticle excitation component relevant to the spin resonance in the multiband system, we introduce the following quantity:
\begin{align}
 \mathcal{L}(\mathbf{k},\mathbf{q})
 &= \sum_{j,l}
 \mathcal{M}^{j,l}_{f}(\mathbf{k},\mathbf{q}) \notag \\
 &\quad \times \left[
 \theta\left(E_j^{\mathbf{k}+\mathbf{q}}\right) \theta\left(-E_l^{\mathbf{k}}\right)
 + \theta\left(-E_j^{\mathbf{k}+\mathbf{q}}\right) \theta\left(E_l^{\mathbf{k}}\right)
 \right].
\label{eq:L_factor}
\end{align}
Here, the Heaviside function $\theta(x)$ singles out quasiparticle
excitations from occupied states to unoccupied states.
Indeed, in the orbital-decoupled limit, $\mathcal{L}(\mathbf{k},\mathbf{q})$ becomes
\begin{equation}
 \mathcal{L}(\mathbf{k},\mathbf{q}) = \frac{1}{2} \left( 1 - \frac{
\varepsilon_f^{\mathbf{k}}
\varepsilon_f^{\mathbf{k}+\mathbf{q}}
+
\Delta^{\mathbf{k}}
\Delta^{\mathbf{k}+\mathbf{q}}
}{
E_{f_+}^{\mathbf{k}}
E_{f_+}^{\mathbf{k}+\mathbf{q}}
} \right),
\end{equation}
which corresponds to the coherence factor in the particle--hole channel [Eq.~\eqref{eq:coherence_factor_f_sector_p-h}].
Therefore, $\mathcal{L}(\mathbf{k},\mathbf{q})$ can be regarded as a
quantity for examining which quasiparticle scattering processes are
enhanced by a sign-changing SC order parameter in the multiband
hybridized system. In this study, we use
$\mathcal{L}(\mathbf{k},\mathbf{q})$ to analyze the origin of the spin
resonance in the SC state.

In the following, we calculate the dynamical spin susceptibility to analyze the correspondence with AFM fluctuations observed in INS experiments. We fix the temperature to $T=0.01$, the mesh size to $N=512\times512\times64$, $\delta=0.001$, and $U=1$.

\section{Results and Discussion}
\label{sec:results}

In this section, we present the magnetic responses in the normal and SC states.
We first analyze how the hybridization between the localized $f$ electrons and conduction electrons modifies the normal-state spin excitation spectrum.
We then examine how the SC order parameter affects the spin response and identify the condition for the superconductivity-induced spin resonance.

\subsection{Magnetic excitation in the normal state}
\label{sec:results_normal}

\begin{figure}[tbp]
  \centering
  \includegraphics[width=\linewidth]{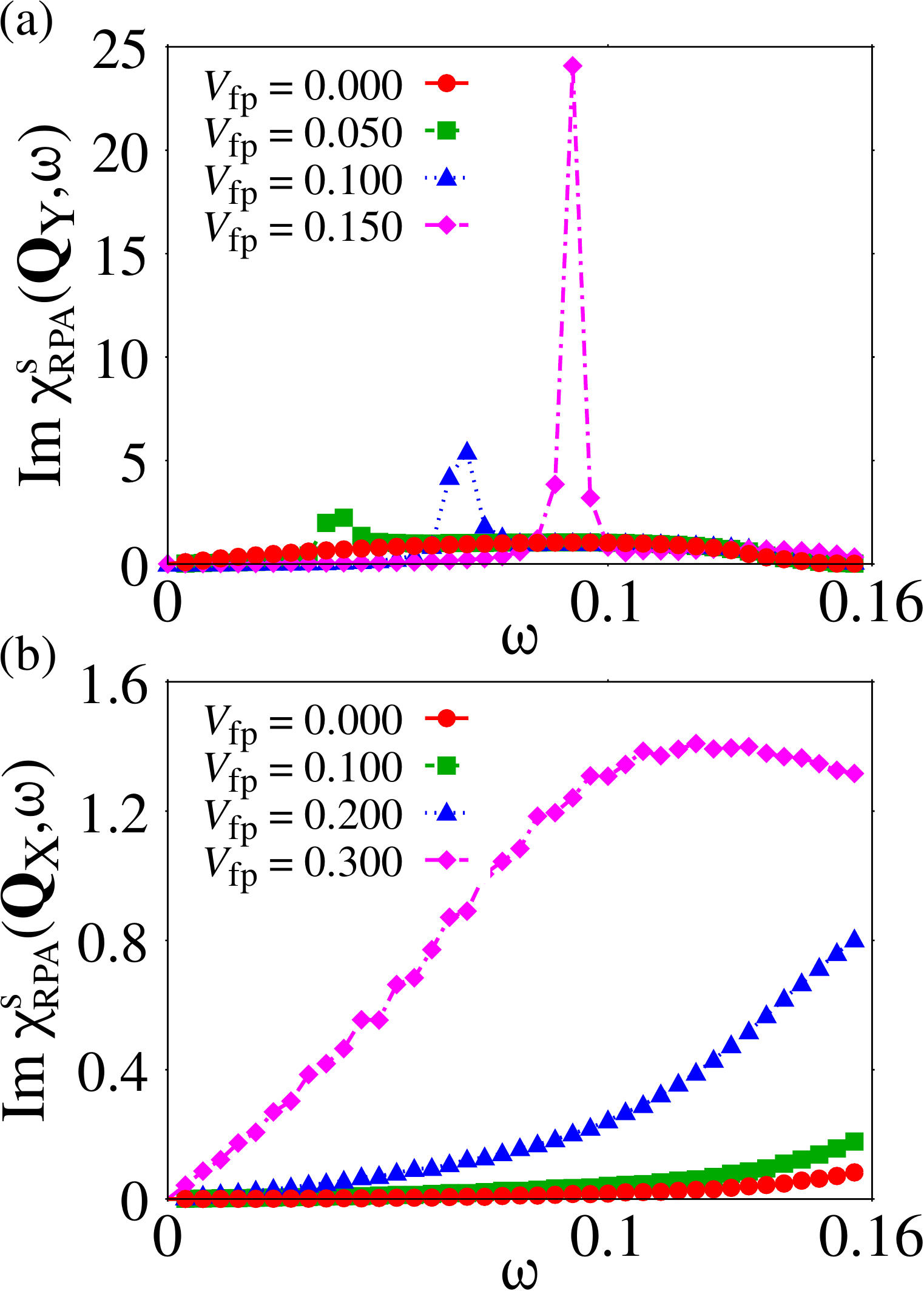}
  \caption{Imaginary part of the RPA spin susceptibility in the normal state plotted as a function of frequency at (a) $\mathbf{q} = \mathbf{Q}_{\mathrm{Y}} = (0,\pi,0)$ and (b) $\mathbf{q} = \mathbf{Q}_{\mathrm{X}} = (\pi,0,0)$. The dependence on the hybridization $V_{fp}$ is shown in both panels.}
  \label{fig:chis_QandQprime}
\end{figure}

\begin{figure}[tbp]
  \centering
  \includegraphics[width=\linewidth]{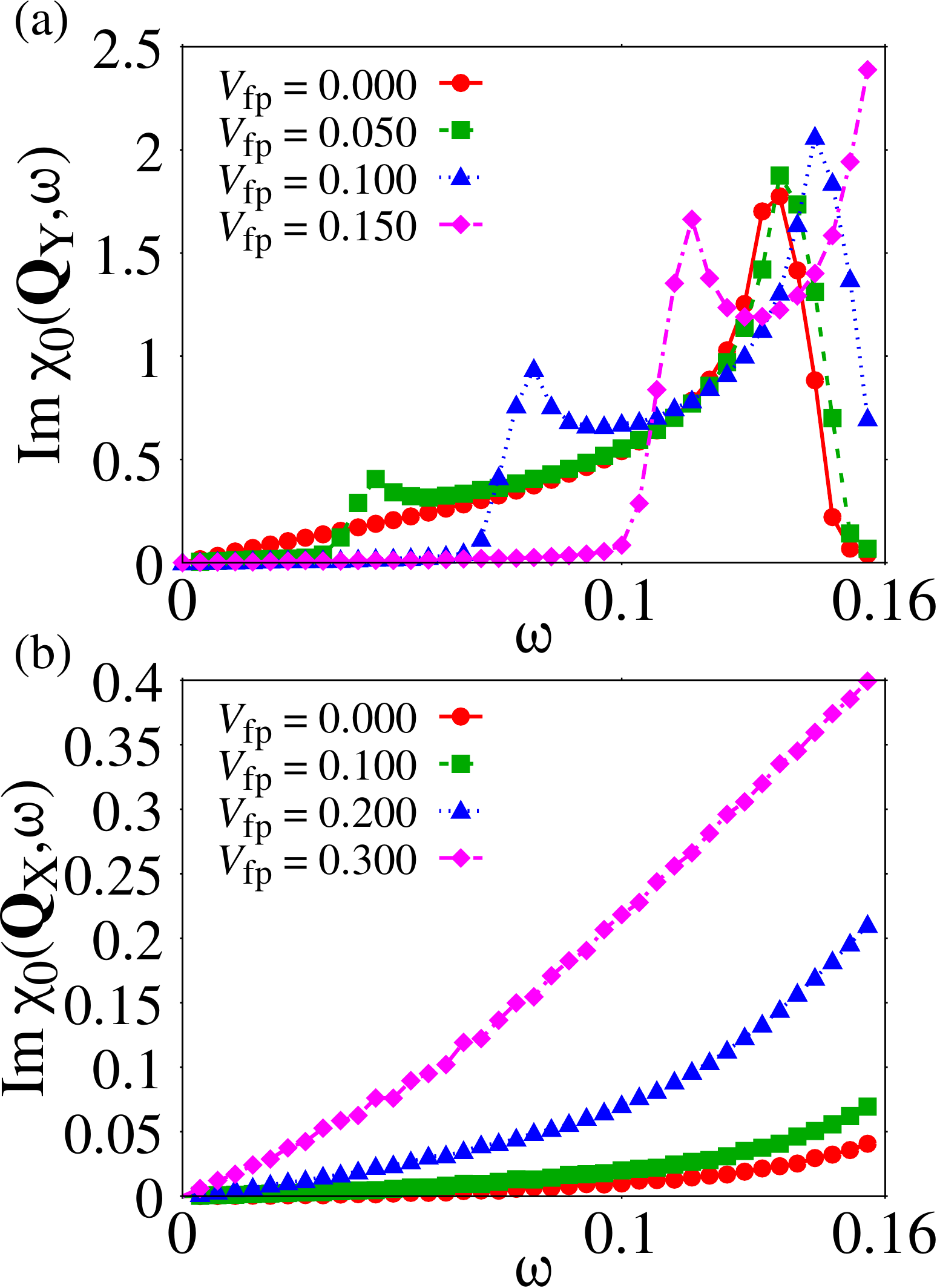}
  \caption{Imaginary part of the bare spin susceptibility in the normal state plotted as a function of frequency at (a) $\mathbf{q} = \mathbf{Q}_{\mathrm{Y}} = (0,\pi,0)$ and (b) $\mathbf{q} = \mathbf{Q}_{\mathrm{X}} = (\pi,0,0)$.}
  \label{fig:chi0_QandQprime}
\end{figure}

We first analyze the imaginary part of the spin susceptibility
$
\Im\chi^{\mathrm{s}}_{\mathrm{RPA}}(\mathbf{Q}_{\mathrm{Y}},\,\omega\bigr)
$
in the normal state at the AFM wave vector $\mathbf{Q}_{\mathrm{Y}}=(0,\pi,0)$. Figure~\ref{fig:chis_QandQprime}(a) shows the imaginary part of the spin susceptibility $\Im\chi^{\mathrm{s}}_{\mathrm{RPA}}(\mathbf{q}, \omega)$ in the normal state for the hybridization $V_{fp} = 0$, $0.05$, $0.1$, and $0.15$. For $V_{fp}=0$, the spectral weight appears continuously from low energies,
and the spectrum does not show a sharp isolated peak. This indicates a
gapless particle--hole continuum and the absence of a collective
excitation. When finite $V_{fp}$ is introduced, the low-energy spectral weight is suppressed, and a pronounced enhancement appears at finite $\omega$. As $V_{fp}$ increases, the spectral peak is enhanced and shifts toward higher energies, suggesting that the enhancement of the spin response is associated with the hybridization between localized $f$ electrons and conduction electrons. 
We interpret this finite-frequency enhancement as a magnetic excitation
associated with the hybridization gap, as discussed previously in Ref.~\cite{Akbari2009}. Our calculations support the experimental interpretation proposed by Halloran et al.~\cite{Halloran2025}, according to which the normal-state magnetic excitation near $\mathbf{q}_\mathrm{exp}=(0, 0.57, 0)\,\mathrm{r.l.u.}$ may originate from an interband spin exciton associated with particle--hole scattering across the hybridization gap between different hybridized quasiparticle bands.

We also examine the spin response at $\mathbf{Q}_{\mathrm{X}}=(\pi,0,0)$ in order to clarify the momentum dependence of the excitation induced by the hybridization $V_{fp}$. 
Figure~\ref{fig:chis_QandQprime}(b) shows the imaginary part of the RPA
spin susceptibility, $\Im\chi^{\mathrm{s}}_{\mathrm{RPA}}(\mathbf{Q}_{\mathrm{X}},\omega)$.
In contrast to the result at $\mathbf{Q}_{\mathrm{Y}}=(0,\pi,0)$, the
spectrum does not exhibit a sharp resonance-like enhancement even when
finite hybridization is introduced.

The difference for the two wave vectors is more clarified by examining the bare spin susceptibility in the normal state. As shown in
Fig.~\ref{fig:chi0_QandQprime}(a), for 
$\mathbf{Q}_{\mathrm{Y}}=(0,\pi,0)$, the low-energy spectral weight of
$\Im\chi_0(\mathbf{Q}_{\mathrm{Y}},\omega)$ is strongly suppressed
when finite hybridization is introduced, and a clear threshold structure
appears at finite energy. In contrast, for
$\mathbf{Q}_{\mathrm{X}}=(\pi,0,0)$, Fig.~\ref{fig:chi0_QandQprime}(b) shows
that $\Im\chi_0(\mathbf{Q}_{\mathrm{X}},\omega)$ remains finite down
to low energies and develops continuously from the low-energy region.
Thus, a hybridization-induced threshold is formed for
$\mathbf{Q}_{\mathrm{Y}}$, whereas no clear threshold structure appears for
$\mathbf{Q}_{\mathrm{X}}$.

This difference in the low-energy structure of $\Im\chi_0(\mathbf{q},\omega)$ leads to
the different responses at the two wave vectors, $\mathbf{Q}_{\mathrm{Y}}$ and $\mathbf{Q}_{\mathrm{X}}$. At $\mathbf{Q}_{\mathrm{Y}}$, the $f$-electron particle--hole continuum extends up to approximately $\omega = 4t_{fy} = 0.14$ for $V_{fp} = 0$, whereas finite hybridization produces a threshold in $\Im\chi_0(\mathbf{Q}_{\mathrm{Y}},\omega)$. This threshold enhances $\Re\chi_0(\mathbf{Q}_{\mathrm{Y}},\omega)$ through the Kramers--Kronig relation and consequently produces a pronounced enhancement of the RPA spin susceptibility. For $\mathbf{Q}_{\mathrm{X}}$, by contrast,
the absence of a clear threshold structure makes this mechanism ineffective,
so that neither $\Re\chi_0(\mathbf{Q}_{\mathrm{X}},\omega)$ nor the
RPA spin susceptibility shows a comparable enhancement.

\begin{figure}[tbp]
  \centering
  \includegraphics[width=\linewidth]{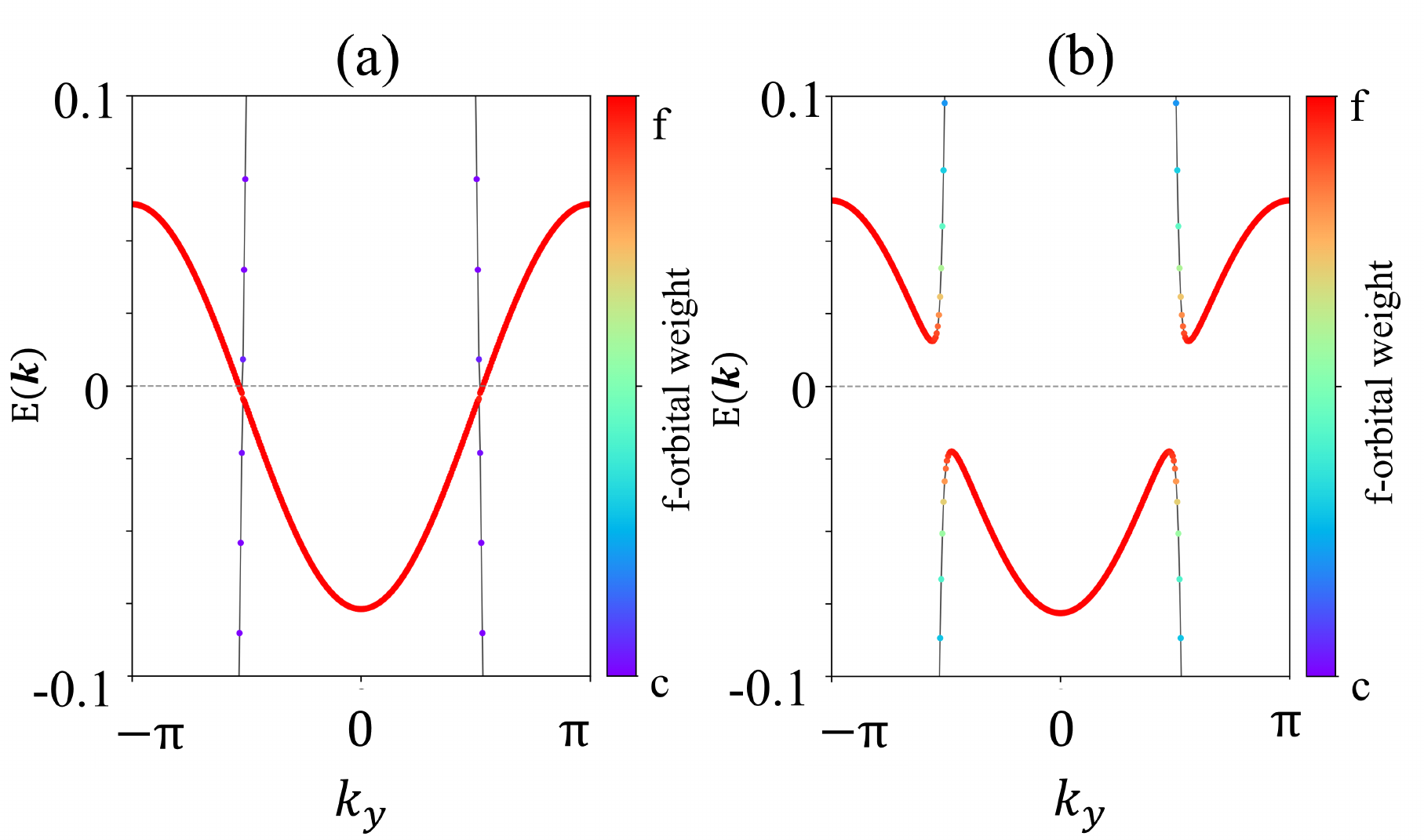}
  \caption{Energy band dispersion $E(\mathbf{k})$ along $(k_x, k_z) = (0, \pi)$ for $-\pi < k_y < \pi$, for (a) $V_{fp} = 0$ and (b) $V_{fp} = 0.05$. The color bars show the orbital weight, with red indicating strong $f$-electron weight and blue indicating strong conduction-electron weight.}
  \label{fig:bandky}
\end{figure}

\begin{figure}[tbp]
  \centering
  \includegraphics[width=\linewidth]{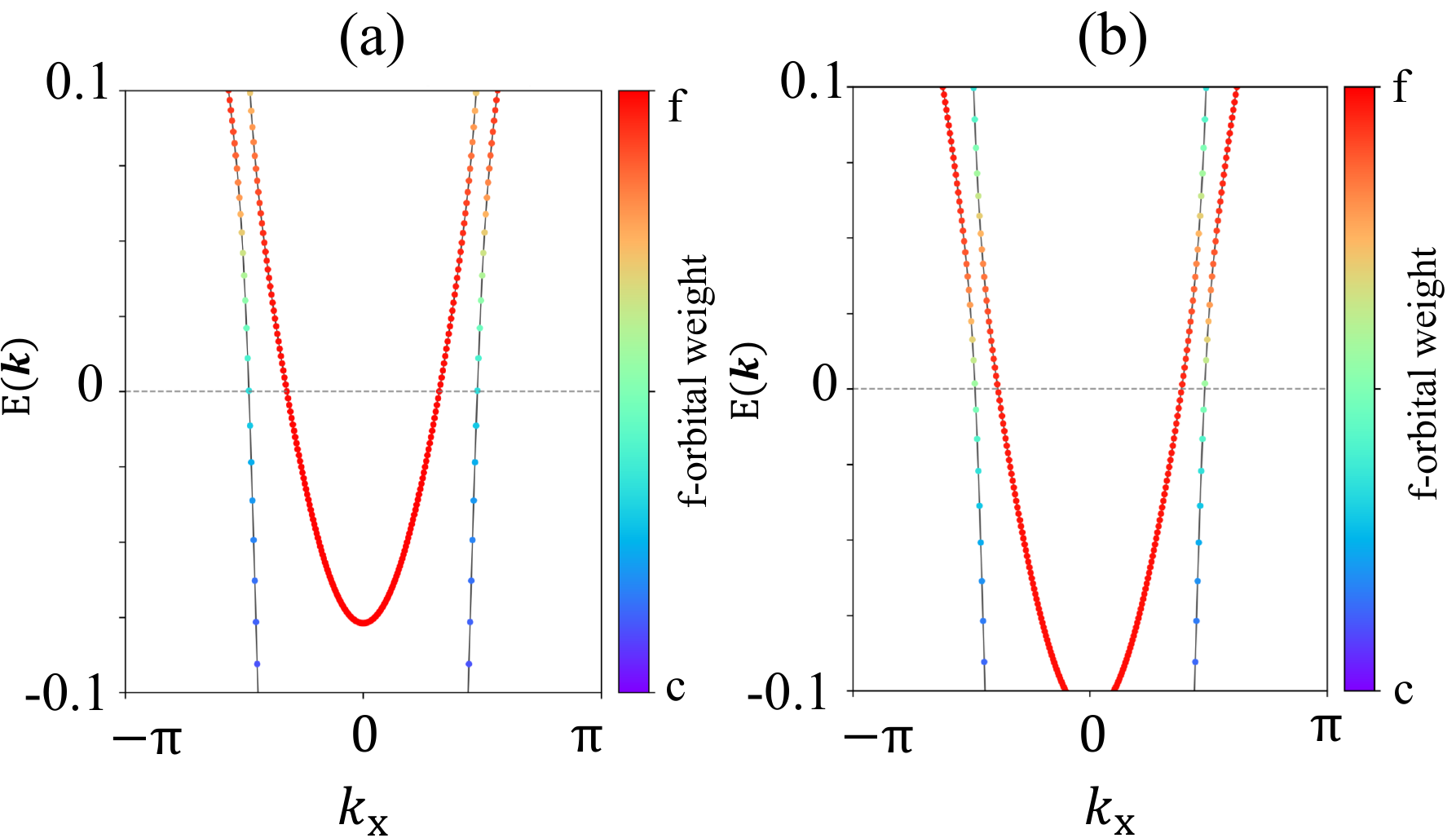}
  \caption{Energy band dispersion $E(\mathbf{k})$ along $(k_y, k_z) = (0, \pi)$ for $-\pi < k_x < \pi$, for (a) $V_{fp} = 0$ and (b) $V_{fp} = 0.26$. The orbital weight is illustrated by the color bar, which is the same as in Fig.~\ref{fig:bandky}.}
  \label{fig:bandkx}
\end{figure}

The presence or absence of a threshold in the normal-state $\Im\chi_0(\mathbf{q},\omega)$ under finite hybridization $V_{fp}$ originates from the energy band dispersion, which is shown in Figs.~\ref{fig:bandky} and \ref{fig:bandkx}. Figure~\ref{fig:bandky} illustrates the low-energy spectrum along the $(k_x, k_z) = (0, \pi)$ line, on which the $f$-electron character is dominant near the Fermi level, for (a) $V_{fp} = 0$ and (b) $V_{fp} = 0.05$. A finite $V_{fp}$ induces an anticrossing near the Fermi level and opens a hybridization gap. However, a hybridization gap does not necessarily appear for all momentum directions. Indeed, Fig.~\ref{fig:bandkx} shows the band dispersion along the $(k_y, k_z) = (0, \pi)$ line, which does not exhibit a gap opening near the Fermi energy even when the hybridization $V_{fp}$ is finite. 

The difference in the band structures between Figs.~\ref{fig:bandky} and \ref{fig:bandkx} can be understood by considering the sign of hopping integrals.
On the $k_z = \pi$ plane, the low-energy spectrum of conduction bands along the $k_y$ ($k_x$) direction is dominated by $p$ ($d$) electrons.
The relevant hopping integral is positive, $t_{py} > 0$, along the $k_y$ direction [see Eq.~\eqref{eq:dispersion_p}], while it is negative ($-t_{dx} < 0$) along the $k_x$ direction [Eq.~\eqref{eq:dispersion_d}].
On the other hand, the in-plane hoppings of $f$ electrons, $-t_{fx}$ and $-t_{fy}$, both have a minus sign [Eq.~\eqref{eq:dispersion_f}], which is opposite to $p$ orbitals but the same as $d$ orbitals.
Therefore, the relative sign difference causes the gapped and gapless excitation spectra in Figs.~\ref{fig:bandky}(b) and \ref{fig:bandkx}(b), respectively.
Note that these simple discussions for the energy bands are possible because of the low-dimensional character of conduction electrons in UTe$_2$.
Indeed, we can qualitatively reproduce the energy band structure in Figs.~\ref{fig:bandky} and \ref{fig:bandkx} and the resultant spin susceptibility in Fig.~\ref{fig:chi0_QandQprime} by using a one-dimensional Anderson model; see Appendix~\ref{app:continuum_onset} for details.

\subsection{Magnetic response in the SC state} 
\label{sec:results_SC}
\begin{figure}[tbp]
 \centering
 \includegraphics[width=0.8\linewidth]{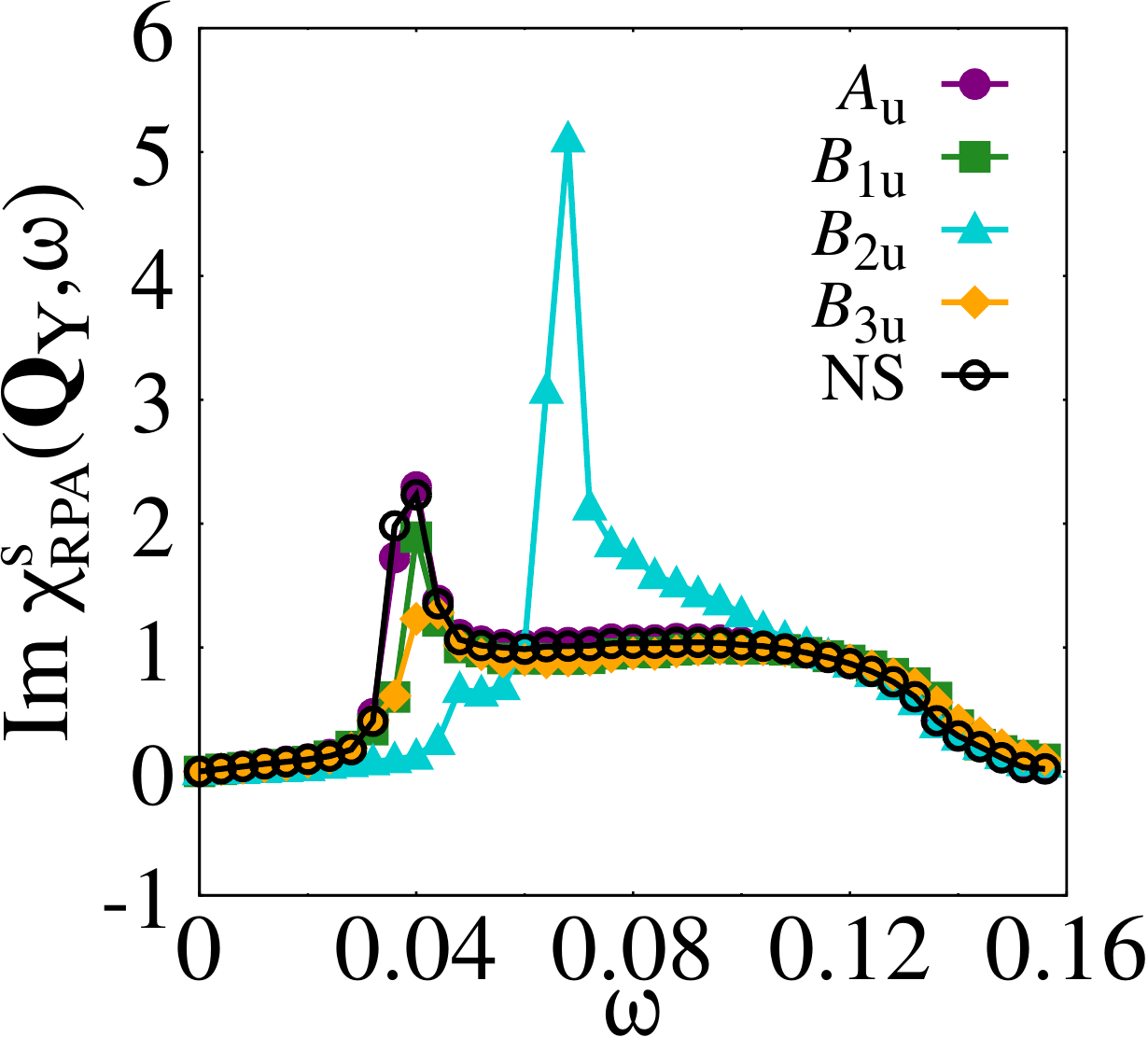}
 \caption{Frequency dependence of the RPA spin susceptibility at $\mathbf{q} = \mathbf{Q}_{\mathrm{Y}} = (0,\pi,0)$ in the normal state (NS) and SC state for each order parameter. Here we fix $\Delta_0 = 0.03$ and $V_{fp} = V_{fd} = 0.05$.}
 \label{fig:chis_QAFM}
\end{figure}

Next, we calculate the dynamical spin susceptibility in the SC state. Figure~\ref{fig:chis_QAFM} shows $\Im\chi^{\mathrm{s}}_{\mathrm{RPA}}(\mathbf{Q}_{\mathrm{Y}},\,\omega)$ in the SC state for all four symmetries in Table~\ref{tab:gap_sym}. Apart from the hybridization-induced spin resonance in the normal state (Sec.~\ref{sec:results_normal}), only the $B_{2u}$ state exhibits a pronounced resonance peak, whereas no additional superconductivity-induced resonance appears for the other three symmetries. Particularly in the $A_u$ case, the spectral shape is almost identical to that in the normal state.

\begin{figure}[t]
 \centering
 \includegraphics[width=0.84\linewidth]{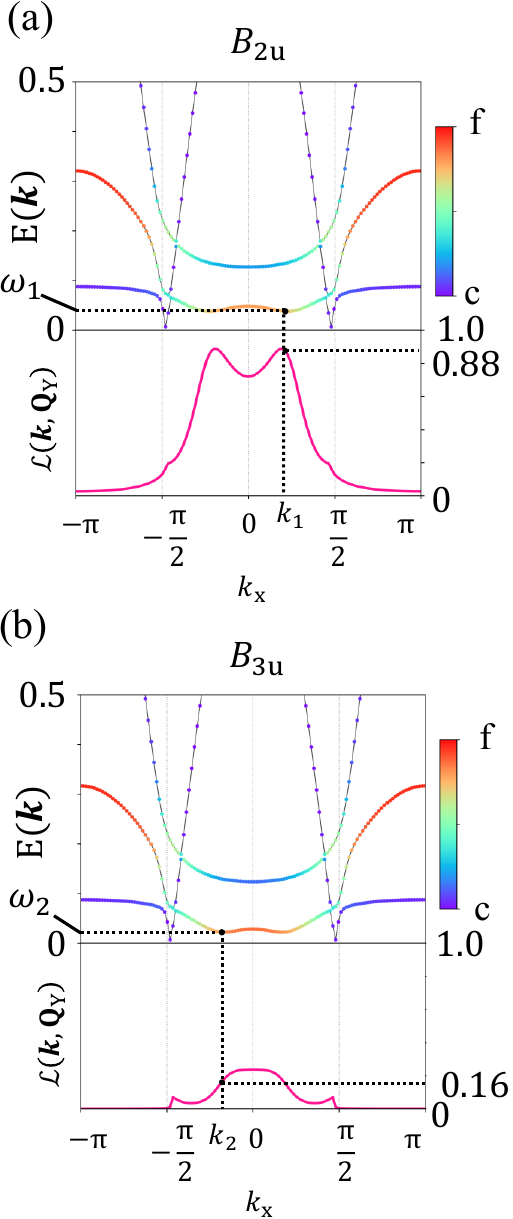}
 \caption{Bogoliubov band dispersions $E(\mathbf{k})$ and coherence factor $\mathcal{L}(\mathbf{k},\mathbf{Q}_{\mathrm{Y}})$ plotted along the $k_x$ direction with $(k_y,k_z)=(-\pi/2,\pi)$. The panels (a) and (b) show the results for the $B_{2u}$ and $B_{3u}$ states, respectively.} In both panels, the upper part shows the Bogoliubov band dispersions, while the lower part shows $\mathcal{L}(\mathbf{k},\mathbf{Q}_{\mathrm{Y}})$. The band dispersions are colored according to the $f$-orbital weight. The left and right axes indicate the values of the energy $E(\mathbf{k})$ and the coherence factor $\mathcal{L}(\mathbf{k},\mathbf{Q}_{\mathrm{Y}})$, respectively.
 \label{fig:coherencefactor}
\end{figure}

To understand why the superconductivity-induced resonance appears only in the $B_{2u}$ state at $\mathbf{Q}_{\mathrm{Y}}$, we inspect $\mathcal{L}(\mathbf{k},\mathbf{Q}_{\mathrm{Y}})$ along a momentum path crossing the Fermi-surface segments connected by the nesting vector $\mathbf{Q}_{\mathrm{Y}}$. As discussed in Sec.~\ref{sec:coherence_factor}, $\mathcal{L}(\mathbf{k},\mathbf{q})$ characterizes the contribution of particle--hole quasiparticle scattering between $\mathbf{k}$ and $\mathbf{k}+\mathbf{q}$ to the bare spin susceptibility in the SC state. This contribution is enhanced by the SC coherence factor when the order parameter changes its sign under the momentum transfer $\mathbf{q}$, resulting in a large $\mathcal{L}(\mathbf{k},\mathbf{q})$ for such sign-reversing scattering processes. Figure~\ref{fig:coherencefactor}(a) shows the Bogoliubov quasiparticle dispersions for $B_{2u}$, together with $\mathcal{L}(\mathbf{k},\mathbf{Q}_{\mathrm{Y}})$ evaluated along a path that crosses Fermi surface segments connected by the nesting vector $\mathbf{Q}_{\mathrm{Y}}$. For comparison, we also plot the same quantities for $B_{3u}$ in Fig.~\ref{fig:coherencefactor}(b). In these figures, the points $\mathbf{k}_1$ and $\mathbf{k}_2$ are the momenta that yield the lowest particle--hole excitation energy in the $B_{2u}$ and $B_{3u}$ states, respectively. Here, we focus on excitations with appreciable $f$-orbital weight, since they provide the dominant contribution to the $f$-electron spin susceptibility. For the $B_{2u}$ order parameter, the lowest excitation occurs at $\mathbf{k}_1$, defining the continuum threshold $\omega_1$ [Fig.~\ref{fig:coherencefactor}(a)]. At this momentum, $\mathcal{L}(\mathbf{k}_1,\mathbf{Q}_{\mathrm{Y}})\simeq 0.88$, indicating that the particle--hole quasiparticle scattering becomes relevant due to the coherence factor. By contrast, in the $B_{3u}$ state, the factor is given by $\mathcal{L}(\mathbf{k}_2,\mathbf{Q}_{\mathrm{Y}})\simeq0.16$ [Fig.~\ref{fig:coherencefactor}(b)], which indicates the suppression of the particle--hole scattering. This behavior reflects the different momentum-space structures of the SC order parameters. The $B_{2u}$ order parameter ($\Delta^{\mathbf{k}}\propto\sin k_y$) changes its sign under the momentum transfer $\mathbf{Q}_Y$, while the $B_{3u}$ order parameter ($\Delta^{\mathbf{k}}\propto\sin k_x$) remains unchanged. In addition, since the relation $\Delta^{\mathbf{k}+\mathbf{Q}_Y}=\Delta^{\mathbf{k}}$ holds in the $B_{1u}$ state ($\Delta^{\mathbf{k}} \propto \sin k_z$), the particle--hole scattering is not enhanced by the SC coherence factor.

\begin{figure}[tbp]
 \centering
 \includegraphics[width=\linewidth]{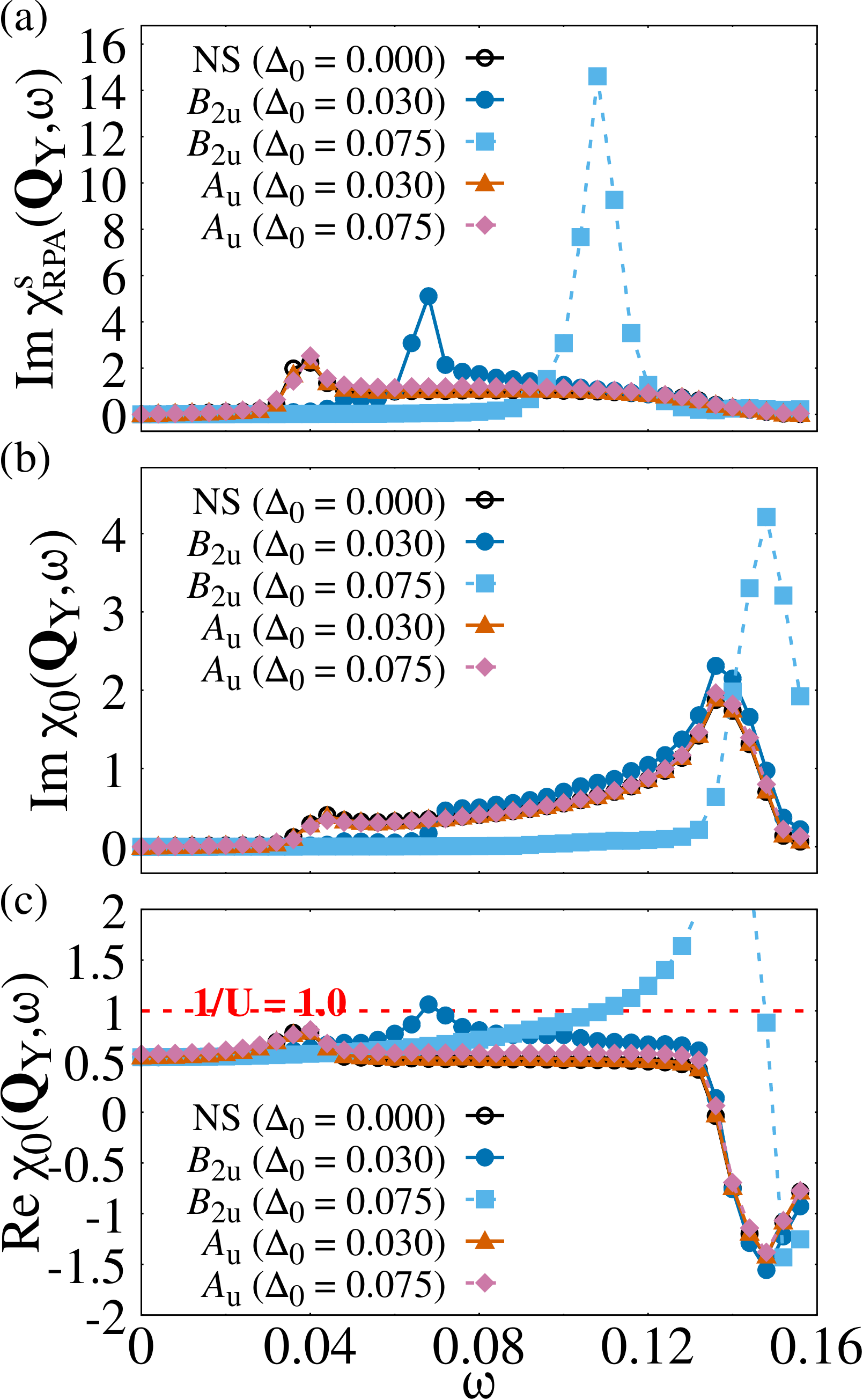}
 \caption{Frequency dependence of the spin susceptibility at $\mathbf{Q}_{\mathrm{Y}}$. Panels (a), (b), and (c) show the imaginary part of the RPA spin susceptibility [$\Im\chi^{\mathrm{s}}_{\mathrm{RPA}}(\mathbf{Q}_{\mathrm{Y}},\omega)$], the imaginary part of the bare spin susceptibility [$\Im\chi_0(\mathbf{Q}_{\mathrm{Y}},\omega)$], and the real part of the bare spin susceptibility [$\Re\chi_0(\mathbf{Q}_{\mathrm{Y}},\omega)$], respectively. In all panels, the normal-state result with $\Delta_0=0$ is compared with the SC states with $B_{2u}$ and $A_u$ pairing symmetries for $\Delta_0=0.03$ and $0.075$. The horizontal dashed line in panel (c) represents $1/U = 1$.}
 \label{fig:Imchis_Rechi0_Imchi0}
\end{figure}

However, no superconductivity-induced spin resonance is observed for the $A_u$ state in Fig.~\ref{fig:chis_QAFM}, even though the order parameter $\Delta^{\mathbf{k}}\propto\sin k_x\sin k_y\sin k_z$ satisfies $\Delta^{\mathbf{k}+\mathbf{Q}_{\mathrm{Y}}}=-\Delta^{\mathbf{k}}$. To clarify the origin of the contrasting behavior between the $A_u$ and $B_{2u}$ states, Fig.~\ref{fig:Imchis_Rechi0_Imchi0}(a) shows the frequency dependence of $\Im\chi^{\mathrm{s}}_{\mathrm{RPA}}(\mathbf{Q}_{\mathrm{Y}},\omega)$ for these two states at different gap amplitudes $\Delta_0 = 0.03$ and $\Delta_0 = 0.075$.
In the $B_{2u}$ state, the RPA spin susceptibility is strongly enhanced as the gap amplitude increases, indicating the appearance of a superconductivity-induced spin resonance.
On the other hand, in the $A_u$ state, the spectral shape remains close to that in the normal state, and no significant enhancement due to superconductivity is observed.

This difference can be understood from the bare susceptibilities shown in Figs.~\ref{fig:Imchis_Rechi0_Imchi0}(b) and \ref{fig:Imchis_Rechi0_Imchi0}(c).
In the $B_{2u}$ state, as shown in Fig.~\ref{fig:Imchis_Rechi0_Imchi0}(b), the SC order parameter suppresses the low-energy spectrum of
$\Im\chi_0(\mathbf{Q}_{\mathrm{Y}},\omega)$,
and produces a clear onset at a finite energy.
From the Kramers--Kronig relation, this leads to an enhancement of
$\Re\chi_0(\mathbf{Q}_{\mathrm{Y}},\omega)$
as shown in Fig.~\ref{fig:Imchis_Rechi0_Imchi0}(c).
As a result, $\Re\chi_0$ approaches the condition for the resonance,
$1-U\Re\chi_0(\mathbf{Q}_{\mathrm{Y}},\omega)=0$,
and the pronounced peak shown in Fig.~\ref{fig:Imchis_Rechi0_Imchi0}(a) appears.
In contrast, in the $A_u$ state, no clear difference from the normal-state results appears for both
$\Im\chi_0(\mathbf{Q}_{\mathrm{Y}},\omega)$
and
$\Re\chi_0(\mathbf{Q}_{\mathrm{Y}},\omega)$.
Although the $A_u$ order parameter changes its sign under $\mathbf{Q}_{\mathrm{Y}}$, the gap function
$\sim \sin k_x\sin k_y\sin k_z$ vanishes on the $k_z=\pi$ plane, where the relevant $f$-electron-dominated Fermi surface is located.
Therefore, the SC gap does not produce a clear additional threshold structure in
$\Im\chi_0$,
nor does it strongly enhance
$\Re\chi_0$.
From these results, the superconductivity-induced spin resonance at $\mathbf{Q}_{\mathrm{Y}}$ appears only in the $B_{2u}$ state.
The peak-like structure observed in the $A_u$ state can be interpreted not as a spin resonance due to the SC feedback effect, but mainly as a reflection of the Kondo-hybridization-gap structure already present in the normal state.

\begin{figure}[tbp]
 \centering
 \includegraphics[width=1.01\linewidth]{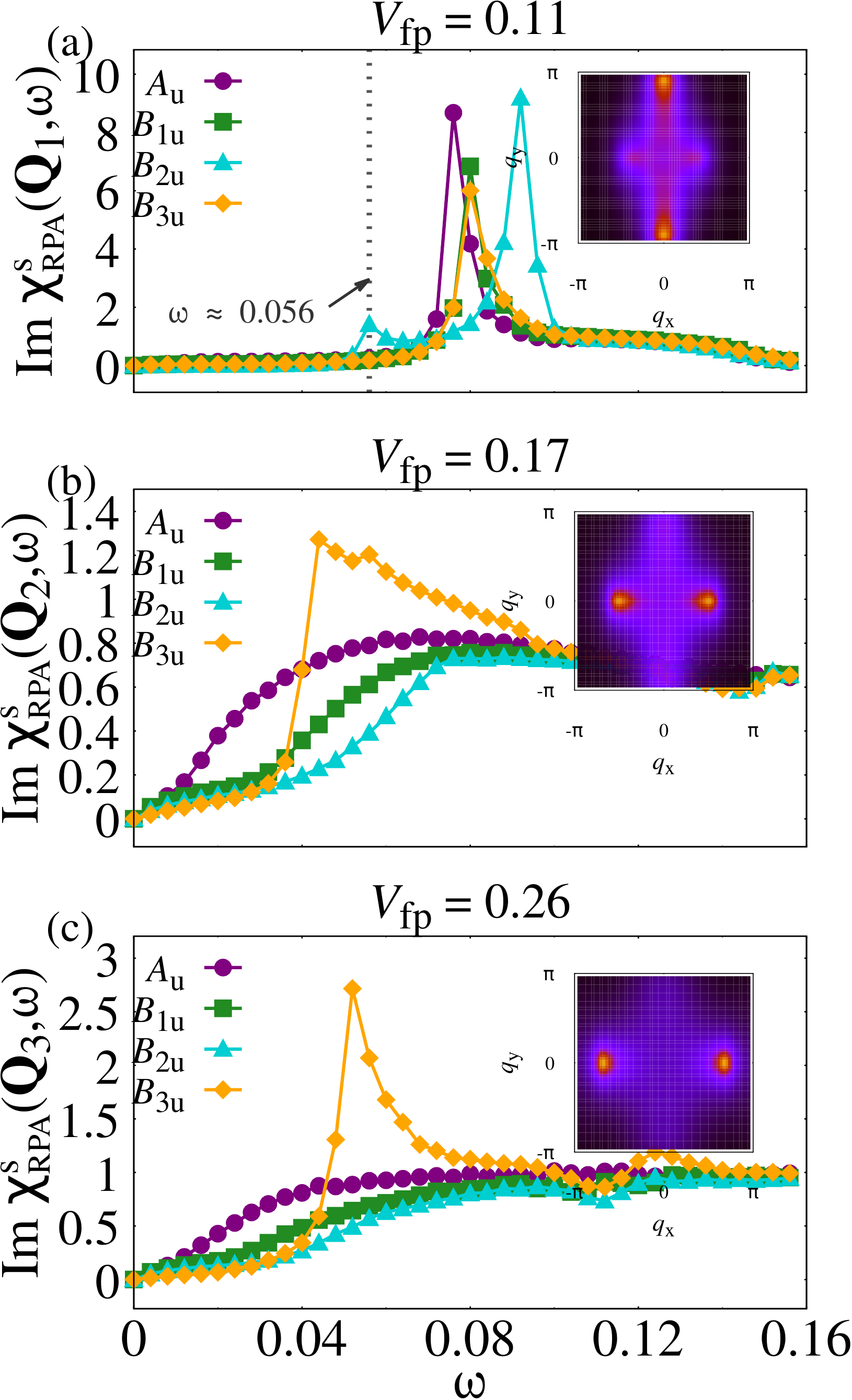}
 \caption{Imaginary part of the dynamical spin susceptibility $\Im\chi^{\mathrm{s}}_{\mathrm{RPA}}(\mathbf{q},\omega)$ calculated for (a) $V_{fp} = 0.11$, (b) $V_{fp} = 0.17$, and (c) $V_{fp} = 0.26$. The wave vector $\mathbf{q}$ is set to the point where the static bare susceptibility $\chi_0(\mathbf{q}, 0)$ reaches the maximum: (a) $\mathbf{Q}_{1}=(0,0.91\pi,0)$, (b) $\mathbf{Q}_{2}=(0.25\pi,0,0)$, and (c) $\mathbf{Q}_{3}=(0.69\pi,0,0)$ (see Fig.~\ref{fig:afm_vectors}). The insets show the color map of the static spin susceptibility on the $q_z=0$ plane; see also Ref.~\cite{Hakuno2024}.}
 \label{fig:eachVfp}
\end{figure}

Finally, we demonstrate that the SC order parameter that enhances the dynamical spin susceptibility $\Im\chi^{\mathrm{s}}_{\mathrm{RPA}}(\mathbf{q},\omega)$ varies as the nesting vector $\mathbf{q}$ changes. In Ref.~\cite{Hakuno2024}, $V_{fp}$ is introduced as a parameter controlling the nesting properties, and it was shown that varying $V_{fp}$ shifts the dominant AFM wave vector from $\mathbf{q} \sim (0,\pi,0)$ to $\mathbf{q} \sim (\pi,0,0)$. Figure~\ref{fig:eachVfp} shows the imaginary part of the dynamical spin susceptibility $\Im\chi^{\mathrm{s}}_{\mathrm{RPA}}(\mathbf{q},\omega)$ calculated with $\Delta_0 = 0.03$ for several values of $V_{fp}$, at the wave vector $\mathbf{q}$ where the static bare spin susceptibility $\chi_0(\mathbf{q},0)$ is maximized.
Figure~\ref{fig:eachVfp}(a) shows $\Im\chi^{\mathrm{s}}_{\mathrm{RPA}}(\mathbf{Q}_1,\omega)$ for $V_{fp}=0.11$, where the static spin susceptibility is maximized at $\mathbf{Q}_1=(0,0.91\pi,0)$. In this case, an enhancement of the spin susceptibility is observed for all SC order parameters. However, as already seen in Fig.~\ref{fig:chis_QandQprime}(a), this enhancement mainly originates from the Kondo hybridization gap and is therefore not a direct manifestation of the SC feedback effect. On the other hand, only for the $B_{2u}$ order parameter does an additional characteristic peak appear around $\omega \approx 0.056 \approx 2\Delta_0$, which is absent for the other SC order parameters. We attribute this peak to the SC feedback effect associated with the sign change $(\Delta_{\mathbf{k}+\mathbf{Q}_1}=-\Delta_{\mathbf{k}})$ for the gap function $(\Delta_{\mathbf{k}}\propto\sin k_y)$.
Figures~\ref{fig:eachVfp}(b) and \ref{fig:eachVfp}(c) show $\Im\chi^{\mathrm{s}}_{\mathrm{RPA}}(\mathbf{q},\omega)$ for $V_{fp}=0.17$ and $0.26$, respectively. For $V_{fp}=0.17$, the maximum of $\chi_0(\mathbf{q},0)$ is located at $\mathbf{Q}_2=(0.25\pi,0,0)$, while for $V_{fp}=0.26$, it is located at $\mathbf{Q}_3=(0.69\pi,0,0)$. Upon changing $V_{fp}$, the dominant nesting vector shifts from being parallel to the $q_y$ direction to being parallel to the $q_x$ direction. At $\mathbf{Q}_2$ and $\mathbf{Q}_3$, none of the four SC states exhibits the pronounced hybridization-induced enhancement, in contrast to the results at $\mathbf{Q}_\mathrm{Y}$. Instead, the $B_{3u}$ state alone exhibits an enhancement induced by superconductivity. This behavior can be understood from the fact that the $B_{3u}$ order parameter has opposite signs on the relevant
$f$-electron-dominated Fermi-surface segments connected by $\mathbf{Q}_2$ and $\mathbf{Q}_3$.

\begin{table*}[t]
  \caption{Summary of the magnetic response for
  dominant momentum transfers parallel to the $q_y$ and $q_x$
  directions. In the SC state, we take $\mathbf{d}(\mathbf{k})\parallel\hat{\mathbf{z}}$ and consider only the longitudinal spin susceptibility. The emergence of a superconductivity-induced
  resonance requires a finite SC gap and a sign reversal of the order
  parameter between the $f$-electron-dominated segments of the Fermi surface connected by the magnetic wave vector.}
\label{tab:resonance_conditions}
  \centering
  \small
  \renewcommand{\arraystretch}{1.25}
    \begin{tabular*}{\textwidth}
    {@{\extracolsep{\fill}}lccccc@{}}
    \hline\hline
    & \multicolumn{1}{c}{Hybridization-induced enhancement}
    & \multicolumn{4}{c}{Superconductivity-induced enhancement} \\
    \cline{2-6}
    Momentum transfer
      & Normal state 
      & $A_u$ & $B_{1u}$ & $B_{2u}$ & $B_{3u}$ \\
    \hline
    $\mathbf{q}\parallel\hat{\mathbf{y}}$
      & \textbf{Yes}
      & No & No & \textbf{Yes} & No \\
    $\mathbf{q}\parallel\hat{\mathbf{x}}$
      & No
      & No & No & No & \textbf{Yes} \\
    \hline\hline
  \end{tabular*}
\end{table*}

Table~\ref{tab:resonance_conditions} summarizes all results for magnetic resonance obtained in our calculations and shows the two distinct resonance mechanisms between the normal and SC states.
The enhancement in the normal state is governed by a hybridization-induced threshold in the particle--hole continuum and hence by the underlying band dispersion.  The superconductivity-induced resonance is instead governed by the SC gap structure. The SC order parameter must remain finite on the relevant $f$-electron-dominated Fermi-surface segments and must change its sign between the quasiparticle states connected by the dominant magnetic wave vector. The change from the $B_{2u}$ response near $\mathbf{Q}_{\mathrm{Y}}$ to the $B_{3u}$ response for momentum transfer along $q_x$ is a direct consequence of this distinction.

\section{Summary}
\label{sec:summary}

In this study, we analyzed the dynamical spin susceptibility in UTe$_2$ based on a mixed-dimensional periodic Anderson model using the BCS-RPA formalism. In particular, we focused on the effect of the hybridization between localized $f$ electrons and conduction electrons on the magnetic response in the normal state, and discuss the effect of the $f$-orbital SC order parameter on the spin resonance in the SC state.

In the normal state, we showed that, when finite hybridization produces a gap structure in the particle--hole continuum, the real part of the bare spin susceptibility is enhanced, leading to an enhancement of the dynamical spin susceptibility through the RPA effect. In particular, a pronounced enhancement appears at $\mathbf{Q}_{\mathrm{Y}}=(0,\pi,0)$, which can be understood as a magnetic excitation originating from the Kondo hybridization gap. On the other hand, this hybridization-induced enhancement is not determined only by the magnitude of the hybridization, but also depends on the direction of the scattering wave vector. Indeed, for momentum transfer along the $q_x$ direction, even in the presence of finite hybridization, no clear gap structure is formed, and the corresponding enhancement of the spin response is suppressed.

In the SC state, taking $\mathbf{d}(\mathbf{k})\parallel\hat{\mathbf{z}}$, we compared the longitudinal spin responses for the odd-parity spin-triplet order parameters $A_u$, $B_{1u}$, $B_{2u}$, and $B_{3u}$. At $\mathbf{Q}_{\mathrm{Y}}=(0,\pi,0)$, a pronounced superconductivity-induced spin resonance appears only in the $B_{2u}$ state. This is because the $B_{2u}$ gap remains finite on the $f$-electron-dominated Fermi surface near $k_z=\pi$ and changes its sign between the quasiparticle states connected by $\mathbf{Q}_{\mathrm{Y}}$. On the other hand, although the $A_u$ state has a sign reversal, the gap vanishes on the relevant Fermi-surface region. Therefore the peak-like structure in the $A_u$ state is attributed not to superconductivity, but to the hybridization-gap structure in the normal state.

Furthermore, when the dominant magnetic wave vector shifts toward the $q_x$ direction, the superconductivity-induced enhancement appears in the $B_{3u}$ state. This result shows that the enhancement of the magnetic response in the normal state and the emergence mechanism of the spin resonance in the SC state are different. The former is governed by whether hybridization forms a finite-energy threshold structure in the particle--hole continuum, whereas the latter is determined by whether the SC gap is finite and changes its sign on the $f$-electron-dominated quasiparticle states connected by the dominant magnetic wave vector. Therefore, to understand magnetic excitations in heavy-fermion multiband superconductors, it is important to distinguish the response originating from the hybridized band structure in the normal state from the response originating from the SC feedback effect.

\begin{acknowledgments}
The authors are grateful to R.~Hakuno, S.~Haruna, and Y.~Yanase for valuable comments.
This work was supported by JSPS KAKENHI Grant No. JP26H00636.
Numerical calculations in this work were partly performed using the facilities of the Supercomputer Center, the Institute for Solid State Physics, the University of Tokyo.
\end{acknowledgments}

\appendix

\section{Continuum onset in a one-dimensional Anderson model}
\label{app:continuum_onset}

In this Appendix, we discuss a minimal mechanism by which the hybridization gap produces different normal-state magnetic responses depending on the momentum direction. 
We consider a one-dimensional Anderson model composed of a conduction band
and an $f$-electron band. Here, we neglect the Coulomb interaction among $f$ electrons.
The Hamiltonian is given by
\begin{equation}
H = \sum_{k}
\begin{pmatrix}
c_k^\dagger & f_k^\dagger
\end{pmatrix}
\hat{H}(k)
\begin{pmatrix}
c_k \\
f_k
\end{pmatrix},
\end{equation}
where $c_k^\dagger$ and $f_k^\dagger$ are the creation operators of conduction and $f$ electrons carrying momentum $k$, respectively.
The Hamiltonian matrix is written as
\begin{equation}
\hat{H}(k) =
\begin{pmatrix}
\varepsilon_c(k) & V \\
V & \varepsilon_f(k)
\end{pmatrix},
\end{equation}
where $V$ is a momentum-independent hybridization.
We take
\begin{equation}
\varepsilon_c(k)=-2s t_c \cos k,\qquad
\varepsilon_f(k)=-2t_f \cos k,
\end{equation}
with $s=\pm 1$ and $0 < t_f < t_c$. We set the chemical potential to zero, corresponding to half filling
in the present particle--hole-symmetric model.
The parameter $s$ represents the relative sign of the effective hopping amplitudes of the two bands. 
The hybridized bands are given by
\begin{equation}
E_{\pm}(k)
=
\frac{\varepsilon_c(k)+\varepsilon_f(k)}{2}
\pm
\sqrt{
\left[
\frac{\varepsilon_c(k)-\varepsilon_f(k)}{2}
\right]^2
+V^2
}.
\end{equation}

We focus on the lower edge of the interband particle--hole continuum at $Q=\pi$. 
At low temperature, it is estimated as
\begin{equation}
\omega
=
\min_k
\left[
E_+(k+Q)-E_-(k)
\right].
\end{equation}
Here, we consider only transitions from occupied to unoccupied states, satisfying $E_-(k)<0$ and $E_+(k+Q)>0$. Introducing $x=\cos k$, the excitation energy can be written as
\begin{equation}
\Delta(x)
=
2(t_f+s t_c)x
+
2\sqrt{(t_f-s t_c)^2x^2+V^2}.
\end{equation}

For $s=+1$, corresponding to hoppings with the same sign, we obtain
\begin{equation}
\Delta_+(x)
=
2(t_c+t_f)x
+
2\sqrt{(t_c-t_f)^2x^2+V^2}.
\end{equation}
Since $\Delta_+(x)$ is a monotonically increasing function on
$-1\leq x\leq1$, its unconstrained minimum occurs at $x=-1$.
After imposing the restriction to transitions from occupied to
unoccupied states, however, the continuum threshold is
\begin{equation}
\Delta_+
=
\max\left\{
0,\,
2\left[
\sqrt{(t_c-t_f)^2+V^2}-(t_c+t_f)
\right]
\right\}.
\end{equation}
Therefore, a finite continuum onset appears only when $V^2>4t_c t_f$ is satisfied.
If $V^2\leq 4t_c t_f$, the interband continuum starts from zero energy.

For $s=-1$, corresponding to hoppings with opposite signs, the excitation energy becomes
\begin{equation}
\Delta_-(x)
=
2(t_f-t_c)x
+
2\sqrt{(t_c+t_f)^2x^2+V^2}.
\end{equation}
In this case, the threshold energy is
\begin{equation}
\Delta_{-}
=
\begin{cases}
\dfrac{4V\sqrt{t_c t_f}}{t_c+t_f},
& \eta \leq 1,
\\[6pt]
2\left[
\sqrt{(t_c+t_f)^2+V^2}
-\lvert t_c-t_f\rvert
\right],
& \eta > 1.
\end{cases}
\label{eq:threshold-opposite}
\end{equation}
Here we introduced the dimensionless parameter
\begin{equation}
\eta
=
\frac{
 V\lvert t_c-t_f\rvert
}{
 2(t_c+t_f)\sqrt{t_c t_f}
}.
\end{equation}
Thus, opposite signs of the effective hoppings naturally generate a finite threshold in the interband continuum.

\begin{figure}[tbp]
 \centering
 \includegraphics[width=1.01\linewidth]{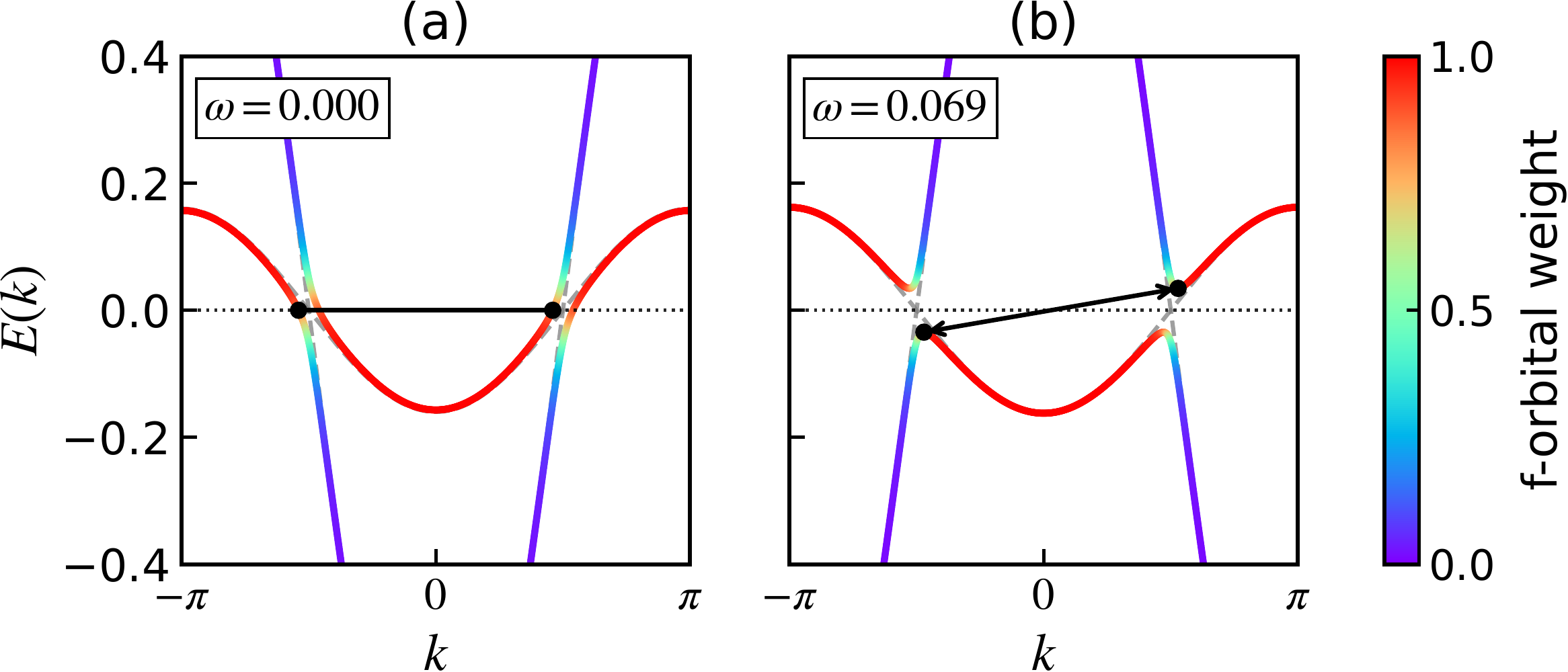}
    \caption{
    Band dispersions of the one-dimensional Anderson model discussed in Appendix~\ref{app:continuum_onset}.
    The dashed gray curves show the unhybridized bands, and the solid curves show the hybridized bands colored by the $f$-orbital weight.
    Panels (a) and (b) correspond to the same-sign and opposite-sign hopping cases, respectively.
    The black points indicate particle--hole excitations connected by $Q=\pi$, and $\omega$ denotes the lower edge of the interband particle--hole continuum.
    }
 \label{fig:appendix_anderson1D}
\end{figure}

The analytic result is illustrated in Fig.~\ref{fig:appendix_anderson1D}.
The representative band dispersions confirm that the lower edge of the
interband particle--hole continuum depends qualitatively on the relative sign
of the effective hopping amplitudes.
For the same-sign case, the continuum remains gapless for the parameters used
in the figure, whereas for the opposite-sign case a finite continuum onset is
formed by the hybridization.

To see the continuum onset, we calculate the $f$-orbital bare spin susceptibility, 
\begin{align}
\chi_0(Q,\omega)
&=
\frac{1}{N}
\sum_{k,\alpha,\beta}
w_{\alpha}^{f}(k+Q)
w_{\beta}^{f}(k) \notag\\
&\times \frac{n\!\left[E_{\alpha}(k+Q)\right]
-
n\!\left[E_{\beta}(k)\right]}{
\omega-
\left[
E_{\alpha}(k+Q)-E_{\beta}(k)
\right]
+i\delta
},
\end{align}
where $\alpha,\beta=\pm$ denote the hybridized bands, $w_{\alpha}^{f}(k)$ is the $f$-orbital weight of band $\alpha$, and $\delta$ is a phenomenological broadening. We then evaluate its imaginary part,
$\operatorname{Im}\chi_0(Q,\omega)$, to identify the lower edge of
the particle--hole continuum.

\begin{figure}[tbp]
 \centering
 \includegraphics[width=0.93\linewidth]{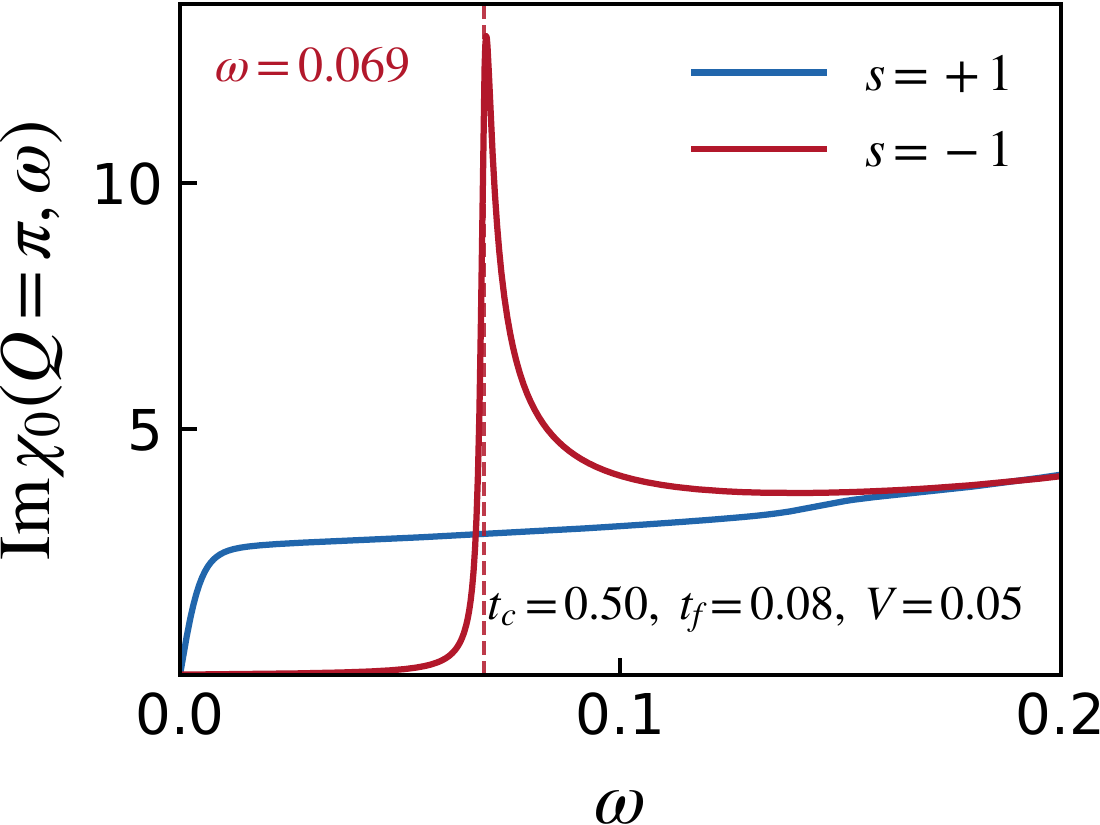}
    \caption{
    Imaginary part of the $f$-orbital bare susceptibility, $\operatorname{Im}\chi_0(Q=\pi,\omega)$, in the one-dimensional Anderson model. The blue and red curves correspond to $s=+1$, for which the conduction and $f$-electron hoppings have the same sign, and (s=-1), for which they have opposite signs, respectively. For $s=+1$, the particle--hole continuum extends to zero energy. For $s=-1$, the spectral weight is suppressed below the finite continuum threshold $\omega=0.069$, indicated by the dashed line. The parameters are $t_c=0.50$, $t_f=0.08$, $V=0.05$, $T=0.001$, and $\delta=0.001$.
    }
 \label{fig:appendix_anderson1D_spectrum}
\end{figure}

Figure~\ref{fig:appendix_anderson1D_spectrum} shows the imaginary part of the $f$-orbital component of the bare spin
susceptibility at $Q=\pi$ for the same-sign ($s=+1$) and
opposite-sign ($s=-1$) hopping cases. The spectral weight extends continuously to zero energy in the same-sign case, whereas it is suppressed below a finite threshold in the opposite-sign case, in agreement with the analytic results derived above.

This result helps us understand the anisotropic normal-state response in the effective model for UTe$_2$. 
The relevant kinetic terms [Eqs.~\eqref{eq:dispersion_f}-~\eqref{eq:dispersion_p}] are
\begin{align}
\varepsilon_f^{\mathbf{k}}
&=
-2t_{fx}\cos k_x
-2t_{fy}\cos k_y
+
2t_{fz}(\cos k_z+1)
+\varepsilon^0_f,\\
\varepsilon_d^{\mathbf{k}}
&=
-2t_{dx}\cos k_x,\\
\varepsilon_p^{\mathbf{k}}
&=
2t_{py}\cos k_y,
\end{align}
where only the hopping terms relevant to the present argument are shown. 
Along the $q_x$ direction, the $f$ and $d$ dispersions correspond to the same-sign case in the above one-dimensional model. 
For $V_{fd}=0.05$, we obtain $V_{fd}^2<4t_{fx}t_{dx}$.
The condition for a finite continuum onset is not satisfied, and the continuum can start from zero energy along the $q_x$ direction. 
This explains why a clear hybridization gap is not necessarily observed in the normal-state spin response along $q_x$, even for relatively large hybridization.

By contrast, along the $q_y$ direction, the $f$ and $p$ dispersions correspond to the opposite-sign case. 
A finite continuum onset is then produced by the hybridization, leading to a gap structure and an enhancement of the spin susceptibility in the normal state near $\mathbf{Q}_{\mathrm{Y}}=(0,\pi,0)$. 
Therefore, the anisotropic normal-state magnetic response originates from the combined effect of the hybridization and the direction-dependent relative sign of the effective hopping amplitudes.

\bibliography{refs}

\end{document}